\documentclass[a4paper,11pt]{article}
\pdfoutput=1 
\usepackage{verbatim}

\usepackage{jheppub}
\usepackage{amsfonts,graphicx,color,amsmath,amssymb,amsfonts,amsxtra, mathrsfs,graphics,graphicx,amsthm,epsfig, youngtab,bm,longtable,float,tikz,empheq}
\usepackage[T1]{fontenc} 
\usepackage{enumerate}
\usepackage{tikz, tikz-3dplot, pgfplots}
\usepackage[utf8]{inputenc}
\usepackage{slashed}
\usepackage{subcaption}

\usepackage{changepage}
\usepackage{float}
\usepackage{tikz}
\usetikzlibrary{decorations.text}
\usepackage{enumitem}
\setlist{itemsep=0pt}
\usetikzlibrary{math}

\usepackage{tikz}
\usepackage{pgfplots}
\usetikzlibrary{decorations.text,intersections, pgfplots.fillbetween}
\usetikzlibrary{math}
 \usetikzlibrary{snakes,3d,shapes.geometric,shadows.blur}
\usepackage{tikz-3dplot}

\makeatletter
\newcommand\footnoteref[1]{\protected@xdef\@thefnmark{\ref{#1}}\@footnotemark}
\makeatother

\newcommand{\comm}[1]{} 

\def\({\left(}
\def\){\right)}
\def\[{\left[}
\def\]{\right]}

\def\One{{\hbox{ 1\kern-.8mm l}}}

\def\barray{\begin{array}}
\def\earray{\end{array}}
\def\be{\begin{equation}}
\def\ee{\end{equation}}
\def\bea{\begin{eqnarray}}
\def\eea{\end{eqnarray}}
\def\bal{\begin{align}}
\def\eal{\end{align}}
\def\nn{\nonumber}

\def\-{\,-\,}
\def\={\,=\,}
\def\+{\,+\,}
\def\equi{\,\equiv\,}

\numberwithin{equation}{section} 

\newcommand{\BR}{\mathbb{R}}

\newcommand{\CM}{{\cal M}}

\newcommand{\wh}{\widehat}

\newcommand{\cM}{\mathcal{M}}

\newcommand{\Secref}[1]{Section~\ref{#1}}

\newcommand{\appref}[1]{App.~\ref{#1}}

\newcommand{\figref}[1]{Fig.~\ref{#1}}
\renewcommand{\eqref}[1]{(\ref{#1})}

\definecolor{cardinal}{rgb}{0.6,0,0}
\definecolor{darkgreen}{rgb}{0,0.4,0}
\definecolor{golden}{rgb}{0.92, 0.7, 0}
\definecolor{midnight}{rgb}{0, 0, 0.5}
\definecolor{darkblue}{rgb}{0, 0, 0.7}
\definecolor{purple}{rgb}{0.5, 0, 0.5}

\def\IR{\mathbb{R}}

\def\cM{{\cal M}}

\def\cQ{{\cal Q}}

\usetikzlibrary{positioning,calc,fadings,decorations.pathreplacing}

\tikzset{
 diffuse color/.initial = black,                       
}

\tikzset{
 linear opacity/.initial=0.5,                          
 linear stroke/.style = {                              
   preaction={                                         
     draw=\pgfkeysvalueof{/tikz/diffuse color},        
     line width = (2.0-#1)*\pgflinewidth,              
     opacity=\pgfkeysvalueof{/tikz/linear opacity},white}},  
 diffuse gradient/.style={                             
   draw = none,                                        
   linear opacity=#1,                                  
   linear stroke/.list={0.0,#1,...,1.0}},              
 diffuse gradient/.default=1,                          
}

\tikzset{
 non-linear stroke/.style = {                          
   preaction={                                         
     draw=\pgfkeysvalueof{/tikz/diffuse color},        
     line width = (2.0-#1)*\pgflinewidth,              
     opacity=#1,white}},                                     
 diffuse falloff/.style={                              
   draw = none,                                        
   non-linear stroke/.list={0.0,#1,...,1.0}},          
 diffuse falloff/.default=1,                           
}
 
\tikzset{%
  >=latex, 
  inner sep=0pt,%
  outer sep=2pt,%
  mark coordinate/.style={inner sep=0pt,outer sep=0pt,minimum size=3pt,
    fill=black,circle}%
}

\title{\boldmath Stability of Topological Solitons, and Black String to Bubble Transition
}

\author[a,b]{Ibrahima Bah,}
\author[a]{Anindya Dey,} 
\author[a]{Pierre Heidmann} 
\affiliation[a]{Department of Physics and Astronomy, Johns Hopkins University,\\
3400 North Charles Street, Baltimore, MD 21218, USA}
\affiliation[b]{Institute for Advanced Study,\\
1 Einstein Drive, Princeton, NJ 08540, USA}

\emailAdd{iboubah@jhu.edu}
\emailAdd{anindya.hepth@gmail.com}
\emailAdd{pheidma1@jhu.edu}

\abstract{We study the existence of smooth topological solitons and black strings as locally-stable saddles of the Euclidean gravitational action of five dimensional Einstein-Maxwell theory.  These objects live in the Kaluza-Klein background of four dimensional Minkowski with an $S^1$.  We compute the off-shell gravitational action in the canonical ensemble with fixed boundary data corresponding to the asymptotic radius of $S^1$, and to the electric and magnetic charges that label the solitons and black strings.  We show that these objects are locally-stable in large sectors of the phase space with varying lifetime.  Furthermore, we determine the globally-stable phases for different regimes of the boundary data, and show that there can be Hawking-Page transitions between the locally-stable phases of the topological solitons and black strings.  This analysis demonstrates the existence of a large family of globally-stable smooth solitonic objects in gravity beyond supersymmetry, and presents a mechanism through which they can arise from the black strings.

}

\preprint{}
\begin{document}

\maketitle
\flushbottom

\newpage 

\section{Introduction}
\label{sec:Intro}

A systematic characterization of the states of quantum gravity and their dynamics continues to be the beast that overshadows most problems of modern physics.  The detection of gravitational waves from colliding ultra compact objects provides important evidence for the existence of astrophysical black holes and novel physical states of gravity whose fundamental nature remains mysterious. This makes the question of quantum gravity even more pressing with potential implications to the emerging field of gravitational-wave spectroscopy.    

Indeed, string theory has provided the best avenue for characterizing the states of quantum gravity and for exploring their dynamics.  This is best exemplified in the accounting of the Bekenstein-Hawking entropy of black holes in many different theories of supergravity from microstates composed of bound states of strings and branes \cite{Strominger:1996sh}.  In a  sense, this accounting exploits open-closed string duality in supersymmetric settings and it allows for the identification of the gravitational states with string and brane states at weak coupling.  It is an open question as to what are the appropriate degrees of freedom or variables in the gravitational theory that characterize the physics of quantum gravity.  Even in the supersymmetric settings that include supergravity such a framework is lacking\footnote{AdS/CFT does provide a description of the states in terms of CFT operators for gravity in Anti-de Sitter space.}.
%

Another important result of string theory is the fact that coherent states of strings and branes can undergo a geometric transition and admit descriptions in supergravity as solitonic solutions, where the gravitational degrees of freedom are characterized by non-trivial topological structures supported by electromagnetic charges.  We refer to them as {\it topological solitons}.  These solutions have been responsible for the many explicit realization of the AdS/CFT duality and have provided important insights into quantum gravity in supersymmetric settings.  In particular, the solitonic solutions can be used to describe coherent microstates of supersymmetric black holes in the fuzzball proposal \cite{Mathur:2000ba,Mathur:2005zp}  and in the microstate geometry program \cite{Bena:2004de,Bena:2006kb,Bena:2007kg,Bena:2015bea,Bena:2016ypk,Bena:2017xbt,Heidmann:2017cxt,Bena:2017fvm,Heidmann:2018mtx,Heidmann:2018vky,Heidmann:2019zws,Heidmann:2019xrd}. 

The question that motivates this paper is: can topological solitons exist in general theories of gravity without supersymmetry and do they teach us anything about the phase space of gravity\footnote{These coherent states would cover a small corner of the phase space of quantum gravity.  The generic state is expected to be very quantumly.}?  These states can be considered as new phases of matter that are inherently geometric and can provide some access to quantum gravity.  An important question will be to understand the types of bound states that can be constructed from them, which may then have real world avatars.  Such new geometric phases of matter could be used to characterize aspects of microstates for astrophysically relevant black holes such as Schwarzschild and Kerr and, as excitingly, to construct new classes of ultra compact objects that range from microscopic to macroscopic scales.  
%

There have been significant hurdles to addressing the above question.  Constructing any solitons in general relativity is a very technical problem, as one must contend with the full non-linear Einstein equations.  In addition to this, there are various no-go theorems on when smooth solitonic solutions can exist in theories of gravity \cite{SeriniSoliton,EinsteinSoliton,Einstein:1943ixi,LichSoliton,Breitenlohner:1987dg,Gibbons:2013tqa}.  Even if we are able to overcome these issues and obtain classical solutions, there is a conceptual question as to when can they be considered as coherent states of quantum gravity.  This requires us to demonstrate that the solutions, at least in the semi-classical regime, are smooth meta-stable saddles of the gravitational path integral.  Significant progress has been made recently in addressing the technical problem in \cite{Bah:2020ogh,Bah:2020pdz,Bah:2021owp,Bah:2021rki,Heidmann:2021cms}, where a framework for obtaining large classes of solitonic solutions in gravity has been developed.  For the first time then, one can systematically study the stability conditions for smooth solitons in gravity beyond supersymmetry.  The main goal of this paper is to initiate a program to tackle this conceptual problem.

In this paper, we study the thermodynamic / quantum-mechanical stability for a family of smooth solitonic solutions described in \cite{Bah:2020ogh,Bah:2020pdz}.  These models are solutions to five dimensional Einstein-Maxwell theory in the Kaluza-Klein background, $\mathbb{M}^4\times S^1_y$ -- four dimensional Minkowski with a circle.  The soliton is spherically symmetric and corresponds to an internal locus where the $S^1_y$ shrinks to zero size.  The solution admits a magnetic charge, and this is crucial for its potential stability as first discussed in \cite{Stotyn:2011tv}.  The asymptotic data that labels these solutions comprises of the magnetic charge and the radius of $S^1_y$. The mass of the soliton can be expressed in terms of these quantities as dictated by the regularity conditions at the aforementioned internal locus\footnote{See \cite{Bah:2021owp,Bah:2021rki} for a more detailed description of on the mechanism that allows for topological solitions in gravity.}. We present a brief summary of these solutions in \Secref{sec:review0}.

The basic object of study for our problem is the thermodynamics of a five dimensional Einstein-Maxwell theory in the canonical ensemble, 
which is introduced in \Secref{sec:review}. The ensemble is defined by putting the theory in a spherical cavity in 
$\mathbb{R}^3 \times S^1_\tau \times S^1_y$, with fixed boundary conditions for the charge and the radius of the thermal and the KK-circle.
Euclidean versions of the solutions of \cite{Bah:2020ogh,Bah:2020pdz} appear as saddle points of the gravitational path integral.
 After carefully defining the ensemble, we explicitly compute the off-shell gravitational action/free energy \cite{York:1986it,Whiting:1988qr,Braden:1987ad,Braden:1990hw} which will be our main tool for probing the thermodynamics of the theory. The off-shell action can be used to find an effective potential, 
which on extremization leads to the different thermodynamic phases of the system and encodes information about their local stability.
 
 In \Secref{sec:infbox}, we study this effective potential and the associated thermodynamic phases in a limit where the spherical cavity is infinitely large.
 We address the question of meta-stability of the thermodynamic phases, as well as the Hawking-Page transitions of one locally-stable phase 
 into another. We summarize our findings in \Secref{sec:conclude}, with a few general remarks about related work in progress.
 The classical stability of the black string in a five dimensional Einstein-Maxwell theory, with both magnetic and electric charge, is discussed in 
 \appref{app:GL-TS}. The details of the off-shell gravitational action for this theory is discussed in \appref{app:electric charge}.

\section{Solutions with solitons}
\label{sec:review0}

In this section we review the solutions of interest discussed in \cite{Bah:2020ogh,Bah:2020pdz}.  We consider a five-dimensional spacetime where one of the spacial direction, $y$, corresponds to a compact extra dimension with $2\pi\beta^\infty_y$ periodicity. The framework corresponds to an Einstein-Maxwell theory with an electric three-form flux, $F^{(e)}$, and a magnetic two-form flux, $F^{(m)}$, described by the following action
\begin{equation}
\begin{split}
I &\=-\frac{1}{16 \pi G_{5}} \int_{\mathcal{M}} d^{5} x \sqrt{g}\left(R-\frac{1}{4} F^{(m)}_{a b} F^{(m)\,a b} -\frac{1}{12} F^{(e)}_{a b c} F^{(e)\,a b c}\right),
\end{split}
\label{eq:Action5d}
\end{equation}
where $G_5$ is the five-dimensional Newton's constant. The solutions are given by two parameters $(r_\text{B},r_\text{S})$ defining the loci of coordinate degeneracies, such as
 \begin{align}
ds^2_5&= -\left( 1 - \frac {r_\text{S}}{r} \right) \, dt^2 +\left( 1 - \frac {r_\text{B}}{r} \right)\, dy^2 + \frac{r^2\,dr^2}{\left(r-r_\text{S} \right) \left(r-r_\text{B} \right) } + r^2\left( d\theta^2 + \sin^2 \theta \,d\phi^2\right), 
\label{eq:metric5d} \\
F^{(e)} &= \frac{Q_e}{r^2}\, dr \wedge dt \wedge dy\,,\qquad F^{(m)} = Q_m\,\sin\theta\,d\theta \wedge d\phi\,. \nonumber
\end{align} 
The electric flux, $F^{(e)}$, is sourced by a line charge $Q_e$ along the $y$ circle and $F^{(m)}$ is sourced by a magnetic monopole $Q_m$. The ansatz for the flux are related by hodge duality and therefore they appear in the Einstein equations in the same way. The Einstein equations are solved if the total charge $\cQ$ is related to the internal parameters as follows 
\begin{equation}
  \mathcal{Q}^2 \equi Q_m^2 +Q_e^2   \= 3\, r_\text{S} r_\text{B} \,.
\label{eq:Field5dGen}
\end{equation}

The family of solutions above asymptote to $\mathbb{R}^{1,3} \times$S$^1_{y}$.    By reducing on the circle, the solutions are those of the Einstein-Maxwell-Dilaton theory with the ADM mass, $\mathcal{M}$, 
\begin{equation}
\mathcal{M}= \frac{2 r_{\text{S}} + r_{\text{B}} }{4 G_4}, \qquad G_4= \frac{G_5}{2\pi \beta^\infty_y}.  
\label{eq:ADMmass}
\end{equation}
The radii $r=r_\text{S}$ and $r=r_\text{B}$ are two special loci with a coordinate degeneracy. More precisely, at $r=r_\text{S}$ the timelike Killing vector, $\partial_t$, vanishes, while at $r=r_\text{B}$ the spacelike Killing vector, $\partial_y$, vanishes. Depending on which one comes first there are two classes of objects in these solutions to study:
\begin{align}
\text{Black String}:& \qquad r_{\text{S}} \geq r_{\text{B}}\,, \label{eq:BScond}\\
\text{topological soliton}:& \qquad r_{\text{B}} > r_{\text{S}} \,. \label{eq:TScond}
\end{align}  When the two mass parameters are equal, $r_{\text{B}} = r_{\text{S}}$, there is an extremal black string solution.  

It is important to note the black string and topological soliton\footnote{In previous literature, the topological solitons were refered to as topological stars.} are related by a double Wick rotation or an analytic continuation of the parameters given as 
\begin{equation}
(t,y) \quad \longleftrightarrow \quad (i t, i y) \qquad \mbox{or} \qquad r_{\text{B}} > r_{\text{S}} \quad \longleftrightarrow \quad r_{\text{B}} < r_{\text{S}} .  \label{wickRo}
\end{equation}  This transformation will be useful in various context in studying the two classes of solutions.

\subsection{Black string} 
\label{sec:BSLorentz}

When $r_\text{S} \geq r_\text{B}$, the locus where the timelike Killing vector $\partial_t$ vanishes is part of the spacetime, thereby leading to a horizon. We will see that for $r_\text{S} > r_\text{B}$, the solutions are non-extremal black strings and that $r_\text{S}=r_\text{B}$ corresponds to extremal black strings.

If $r_\text{S} > r_\text{B}$, the first coordinate singularity is a horizon at $r=r_\text{S}$. The topology of the horizon can be made manifest by considering the local metric with the radial direction
\begin{equation}
\rho^2 \equi \frac{4 \,(r-r_\text{S})}{r_\text{S}-r_\text{B}}\,r_\text{S}^2\,,
\end{equation}
and taking $\rho \rightarrow 0$. The five-dimensional metric \eqref{eq:metric5d} leads to
\begin{equation}
ds_5^2 \,\sim \, - \frac{r_\text{S} -r_\text{B}}{4 r_\text{S}^3}\,\rho^2 \,dt^2 \+ d\rho^2 \+ r_\text{S}^2 \,\left( d\theta^2+ \sin^2 \theta \,d\phi^2 \right) + \frac{r_\text{S}-r_\text{B}}{r_\text{S}} \,dy^2\,.
\end{equation}
The horizon has a S$^2\times$S$^1$ topology and the radii of the S$^2$ and S$^1$ are $r_\text{S}$ and $\sqrt{\frac{r_\text{S}-r_\text{B}}{r_\text{S}}}R_y$ respectively. The Bekenstein-Hawking entropy and the inverse temperature $\beta^\infty_t = T^{-1}$ give
\begin{equation}
S \= \frac{\pi}{G_4}\,\sqrt{r_\text{S}^{3} \,(r_\text{S}-r_\text{B})}\,, \qquad \beta^\infty_t \= \frac{2 r_\text{S}^{\frac{3}{2}}}{\sqrt{r_\text{S} -r_\text{B}}}\,.
\label{eq:betatinf}
\end{equation}

When $r_\text{B}$ approaches $r_\text{S}$, we see that the entropy and the temperature goes to zero, thereby leading to an extremal solution. More precisely, for $r_\text{B}=r_\text{S}$ the solutions \eqref{eq:metric5d} give
\begin{align}
ds^2_5 &\=  \left(1+ \frac{r_\text{S}}{\rho} \right)^{-1}\,\left(-dt^2 + dy^2 \right)+ \left(1+ \frac{r_\text{S}}{\rho} \right)^2\left[d\rho^2 + \rho^2 \,d\Omega_2^2 \right]\,,\\
F^{(e)} &\= \frac{Q_e}{r_\text{S}} d\left(\left(1 + \frac{r_\text{S}}{\rho} \right)^{-1}\right) \wedge dt \wedge dy\,,\qquad F^{(m)} = Q_m\,\sin\theta\,d\theta \wedge d\phi\,,  \qquad Q_m^2 +Q_e^2 =3 r_\text{S}^2 \,, \nonumber
\end{align} 
where we have now defined $\rho =r- r_\text{S}$. We recognize a two-charge extremal black string. At $\rho=0$, both $\partial_y$ and $\partial_t$ Killing vectors degenerate defining an AdS$_3\times$S$^2$ near-horizon geometry.

In this paper, we will study the Euclidean action of the system in the canonical ensemble where the charges, the periodicities of the euclidean time and the $y$-circle are kept fixed. It is therefore relevant to reverse the perspective and express the solutions in terms of $\beta^\infty_t$, $\beta_y^\infty$ and the total charge $\cQ$ \eqref{eq:Field5dGen}. For given $(\beta^\infty_t,\beta_y^\infty,\cQ)$, we found two black string solutions $(r_{\text{S} \,\pm},r_{\text{B}\,\pm})$ such as
\begin{equation}
r_{\text{S} \,\pm} \= \frac{\beta^\infty_t}{2\sqrt{2}} \sqrt{1 \pm \sqrt{1-\frac{16 \cQ^2}{3{\beta^\infty_t}^2}}}\,,\qquad r_{\text{B} \,\pm} \= \frac{\cQ}{\sqrt{6}} \sqrt{1 \mp \sqrt{1-\frac{16 \cQ^2}{3{\beta^\infty_t}^2}}}\,.
\label{eq:2BSinfiniteVol}
\end{equation}
The existence of a black string solution requires then
\begin{equation}
4\cQ \leq \sqrt{3}\,\beta^\infty_t\,.
\end{equation}
Moreover, for fixed temperature and charge in this regime, we have two solutions such that the ``$+$'' solution is bigger than the ``$-$'' one, that is $\cM_+ \geq \cM_-$, where $\cM_\pm$ is their ADM mass \eqref{eq:ADMmass}:
\begin{equation}
\cM_\pm \= \frac{\beta_t^\infty}{16 \sqrt{2}  G_4} \left(5 \pm \sqrt{1-\frac{16  \cQ^2}{3{\beta^\infty_t}^2}} \right)\sqrt{1 \mp \sqrt{1-\frac{16\cQ^2}{3{\beta^\infty_t}^2}}}\,.
\end{equation}

\subsection{Topological soliton}

We now assume that $r_\text{B} > r_\text{S}$. Thus, the outermost coordinate singularity corresponds to $r=r_\text{B}$ where the $y$-circle shrinks to zero size forming an end to spacetime. The second coordinate singularity is not part of the spacetime. The solutions are smooth geometries provided that the metric is regular at $r=r_\text{B}$ where the $y$-circle shrinks\footnote{The regularity outside the coordinate singularity, as the absence of closed timelike curves or the degeneracy of the $\phi$-circle at $\theta=0$ and $\pi$, is fairly straightforward from the form of the metric and gauge fields.}. The region near $r=r_\text{B}$ is best described by the local radial direction
\begin{equation}
\rho^2 \equi \frac{4 \,(r-r_\text{B})}{r_\text{B}-r_\text{S}}\,,
\end{equation}
and taking the limit $\rho \rightarrow 0$. The five-dimensional metric \eqref{eq:metric5d} converges to
\begin{equation}
\begin{split}
ds^2_5 & \sim  - \frac{r_\text{B} - r_\text{S}}{r_\text{B}}\,dt^2 \+ r_\text{B}^2 \,\left[d\rho^2 + \frac{r_\text{B}-r_\text{S}}{4 \,r_\text{B}^3}\,\rho^2 \,dy^2+d\theta^2+\sin^2\theta\,d\phi^2\right]  \,.
\end{split}
\label{eq:bubblemetricBP}
\end{equation}
The $(\theta,\phi)$-subspace describes a round S$^2$ of radius $r_\text{B}$ while the $(\rho,y)$-subspace corresponds to an origin of $\IR^2$. We will allow the local metric to have a conical defect and has the topology of a smooth $\mathbb{Z}_k$ quotient over $\IR^2 \times$S$^2$. This constrains the periodicity of the $y$-circle to be
\begin{equation}
k \beta_y^\infty \= \frac{2 r_\text{B}^\frac{3}{2}}{\sqrt{r_\text{B} - r_\text{S}}}\,, \qquad k\in \mathbb{N}.
\label{eq:Ryconstraint}
\end{equation}
With this condition, the topology at the coordinate singularity corresponds to a \emph{bolt}, a S$^2$ bubble sitting at an orbifolded $\IR^2$. Even in the presence of an orbifold singularity for $k>1$, one can consider the spacetime as smooth. Indeed, within string theory, spacetimes with conical singularities can be smoothed, and often describe localized objects. 

The conical deficit just amplifies the effect of the periodicity of the $y$ circle on the parameters $(r_\text{S},r_\text{B})$. One can for instance consider the rescaled periodicity
\begin{equation}
\widetilde{\beta}_y^\infty \equi k\,\beta_y^\infty,
\end{equation}
and all quantities will depend on $\widetilde{\beta}_y^\infty$ only.

As for the black string, we aim to consider the time and $y$ periodicities and the charges as fixed quantities. For given $(\beta_y^\infty,\beta_t^\infty,\cQ)$ we have a tower of pairs of topological soliton solutions labeled by their orbifold parameter $k$, $(r^{(k)}_{\text{S} \,\pm},r^{(k)}_{\text{B} \,\pm}) $, such as
\begin{equation}
r^{(k)}_{\text{B} \,\pm} \= \frac{k\,\beta^\infty_y}{4\sqrt{2}\,\pi} \sqrt{1 \pm \sqrt{1-\frac{16 \cQ^2}{3k^2\,{\beta^\infty_y}^2}}}\,,\qquad r^{(k)}_{\text{S} \,\pm} \= \frac{\cQ}{\sqrt{6}} \sqrt{1 \mp \sqrt{1-\frac{16 \cQ^2}{3k^2\,{\beta^\infty_y}^2}}}\,.
\end{equation}
There are no restrictions on the existence of a topological soliton geometry for given $(\beta_y^\infty,\beta_t^\infty,\cQ)$, since there always exists a $k$ where the above quantities start to be well-defined. As for the black string, the ``$+$'' solutions can be considered as the bigger solutions compared to the ``$-$'' ones, since the former has a larger radius and ADM mass \eqref{eq:ADMmass}, i.e. $\cM^{(k)}_{+} > \cM^{(k)}_{-}$, where the ADM masses 
$\cM^{(k)}_\pm$ are given as:
\begin{equation}
\cM^{(k)}_\pm \= \frac{k\,\beta_y^\infty}{16 \sqrt{2}  G_4} \left(5 \pm \sqrt{1-\frac{16 \cQ^2}{3k^2{\beta^\infty_y}^2}} \right)\sqrt{1 \mp \sqrt{1-\frac{16 \cQ^2}{3k^2{\beta^\infty_y}^2}}}\,.
\end{equation}

\subsection{Classical Stability}\label{sec:CS}

Before we proceed, we comment on the classical stability of the solution.  It is well known that neutral black strings can suffer from a Gregory-Laflamme instability \cite{Gregory:1993vy}.  Moreover, some magnetic black strings, that are different from the one discussed here, have also been found to be classically unstable \cite{Gregory:1994bj}. It is reasonable to ask when those instabilities exist for the family of solutions considered in this paper.

The stability of the black string solutions \eqref{eq:BScond} without electric charge was studied in \cite{Miyamoto:2006nd}. It was argued that there are no Gregory-Laflamme (GL) instabilities when $ \frac{1}{2} r_{\text{S}}   <  r_{\text{B}} \leq r_\text{S}$. We extend the computation to include an electric charge 
in \appref{app:GL-TS} and show that the above condition for stability still holds. We can use this result to comment on the classical stability of the topological soliton by exploiting the double Wick rotation in \eqref{wickRo}\footnote{This correspondence between the classical instabilities of a bubble and a black string was also pointed out in \cite{Sarbach:2004rm, Stotyn:2011tv}.}.  


The GL instability of the black string can be established by obtaining the static linear deformation mode that sits at the onset of the instability.  In \appref{app:GL-TS}, the ansatz for this mode has zero frequency and generic $y$-momentum: $(\omega =0, k)$.  Under the Wick rotation of \eqref{wickRo} we can track this static mode and find that it becomes a {\it time-dependent} solution of the linear perturbation around the topological soliton.  It exists for the topological soliton when $r_{\text{B}} > 2 r_{\text{S}} $ with frequency and $y$-momentum given as $(i k , 0)$.  The static mode of the black string has a $\cos(ky)$ dependence which maps to a $\cosh( k t)$ time-dependence for the topological soliton.  This signals classical instability for the topological soliton when $r_{\text{B}} > 2 r_{\text{S}} $.  The double Wick rotation of the static black string mode includes the GL threshold mode for the topological soliton with zero angular momentum along the $y$-circle, i.e. the $(\omega =0, k=0)$ mode.  It would be interesting to obtain the general threshold mode for arbitrary momentum.  

This discussion so far has considered instabilities for the black string that can be used to identify some instability of the topological soliton.  We have not said anything about topological solitons in the range $0<r_{\text{S}} < r_{\text{B}} < 2 r_{\text{S}} $.  Suppose there were GL instabilities for the topological soliton in this range; the double Wick rotation would then map them to time-dependent modes for the black string in the range $0<r_{\text{B}} < r_{\text{S}} < 2 r_{\text{B}} $.  However, the black string does not have any linear instabilities in this range.  Thus, we do not expect any GL instability for the topological soliton when $0<r_{\text{S}} < r_{\text{B}} < 2 r_{\text{S}} $.  This could be further checked by looking for the GL threshold modes in general.

%
%

\section{Free energy and potential in the canonical ensemble}
\label{sec:review}

In this section, we carefully study the thermodynamics of a five-dimensional Einstein-Maxwell theory in the canonical ensemble. 
To begin with, let us summarize some of the relevant generalities on thermodynamic ensembles and phase transitions 
in the context of semi-classical gravity. Readers familiar with this material can move directly to \Secref{sec:saddle}.

The starting point is the theory of Euclidean Einstein gravity with a boundary, possibly with other fields in the 
bulk, given by the classical action:
\be 
I [g,\Phi]= I_{\rm gravity}[g] + I_{\rm extra}[g,\Phi], \quad I_{\rm gravity} = I_{\rm EH}+ I_{\rm GH}, 
\ee
where $I_{\rm gravity}$ is the purely gravitational term consisting of the Einstein-Hilbert action and the Gibbons-Hawking boundary term, 
while $I_{\rm extra}$ denotes the action of all the other fields in the gravitational background. Note that $I_{\rm extra}$ may also include 
additional boundary terms, which might be necessary to ensure that the variation of action vanishes at the boundary for the classical 
field configurations. A complete description of the theory requires specification of boundary conditions and  
this is equivalent to specifying the particular thermodynamic ensemble that one is working in. 
Given the theory and the ensemble, one can define the quantum thermal partition function in the standard fashion:
\be
Z= \int [dg][D\Phi] e^{- I[g, \Phi] + I_{\rm gf}},
\ee
where $I_{\rm gf}$ is an appropriate gauge fixing term. For the canonical ensemble,
we fix the size of the bulk (i.e. the position of the boundary), the radius of the thermal circle and the sizes of other compact directions at the 
boundary, as well as the total charge (if any) enclosed by the boundary. In the presence of 
electric fields, it is also possible to define a grand-canonical ensemble, where one fixes an electric potential on the boundary instead of
the charge. For the canonical ensemble (and similarly in the grand-canonical), the thermodynamic quantities can be extracted from 
the thermal partition function in the standard way:
\be
Z= \int [dg][D\Phi] e^{- I[g, \Phi] + I_{\rm gf}} =: e^{-\beta F}, \qquad F = E - T\,S,
\ee
where $F$ is the free energy. We will treat this system in the semi-classical or saddle-point approximation. In this regime, 
the physical quantities of interest can be computed by small quantum fluctuations about the 
classical saddle-points, i.e. solutions of the classical equations of motion consistent with the boundary conditions. 
In close analogy to the semi-classical problem in Quantum Mechanics, the stability of a saddle-point can be inferred by studying 
eigenmodes of the associated Lichnerowicz operator. Presence of negative eigenmodes of the Lichnerowicz operator 
implies that the saddle-point in question is not a minimum of the semi-classical theory, which is often phrased 
as the saddle point being \textit{locally unstable}. 
Locally unstable saddle-points are also referred to as gravitational instantons, and we will say more about them momentarily. 
Absence of any negative eigenmode, therefore implies that the saddle point is locally stable and is a minimum of the theory, at least in the 
semi-classical regime. \\

A thermodynamic phase, in the semi-classical regime, is defined as a saddle-point of the theory.
A thermodynamic phase is \textit{locally stable} if the saddle-point is locally stable in the sense described above.
The thermodynamic ensembles are well-defined for these phases, where quantities like specific heat are positive. 
Given a collection of locally stable thermodynamic phases, a globally stable phase is the one which has the minimum free energy. 
For a set of locally stable saddle-points, one can therefore encounter one of the two following scenarios:
\begin{itemize}

\item A locally stable phase can undergo a \textbf{Hawking-Page transition} \cite{Hawking:1982dh} to the globally stable phase. 
This transition is mediated by a gravitational instanton, 
defined above, where the gravitational instanton interpolates between the locally stable phase and the nucleated phase. The nucleated phase is obtained by an 
appropriate analytic continuation of the graviatational instanton to a Lorentzian space-time. The nucleated phase is thermodynamically 
unstable, by definition, and can evolve semi-classically to reach the globally stable phase. The decay rate of the locally stable phase, computed in the 
semi-classical regime, is given by:
\be
\Gamma = \Gamma_0\, e^{- \left(I_{\rm grav. inst.} - I_{\rm loc. stab.}\right)},
\label{eq:DecayRateGen}
\ee
where $I_{\rm grav. inst.}$ is the classical action of the gravitational instanton, $I_{\rm loc. stab.}$ is the classical action of the locally stable phase, and $\Gamma_0$ is a dimensionful pre-factor that depends on the boundary quantities. More precisely, $\Gamma_0$ is a polynomial function in terms of the Newton constant, temperature, charge and periodicity of the extra dimension\footnote{The pre-factor is obtained from the computation of one-loop determinants around each instanton that mediates the transition. See \cite{Gross:1982cv} for the four-dimensional Schwarzschild solution.}.

\item  There can be multiple phases with the same free energy, which implies that such thermodynamic phases can coexist. One can have a more standard 
first-order phase transition between two such phases.

\end{itemize}

An efficient tool for studying the locally stable phases of a theory in semi-classical gravity is the \textit{reduced 
action} technique, developed in a set of papers \cite{York:1986it,Whiting:1988qr,Braden:1987ad,Braden:1990hw}. 
Given a type of classical saddle point, the construction 
of the reduced action proceeds as follows. We consider a family of metrics with the following properties:
\begin{itemize}
\item The family has the same group of isometries as the classical saddle point in question. Therefore, a 
static and spherically symmetric saddle point will imply that the family of metrics is given by an ansatz 
which obeys staticity and spherical symmetry. 

\item Any representative of the family obeys compatible regularity conditions. For example, given a black string type saddle 
point, this would imply that each representative metric obeys a regularity condition at the horizon. 


\item The metrics obey the boundary conditions for the given ensemble. 
\end{itemize}

Given such a family of metrics, one can solve the Hamiltonian and momentum constraints\footnote{For details, see any standard textbook like \cite{Hawking:1973uf}.} in the Einstein's 
equations, and use them to rewrite the action. The resultant action, obtained after eliminating the 
constraints, is called the \textit{off-shell reduced action}.  Generally, it will be a function of several variables that label the 
family of metrics satisfying the constraints, boundary conditions and regularity conditions. The free energy 
of the theory in the semi-classical regime is given by:
\be
F = \beta^{-1}\, I,
\ee
where $I$ is the reduced action. The extrema of the action (or equivalently, the free energy) with respect to 
these variables give the precise saddle points of the theory. 
The local stability of the extrema can then be studied in the standard fashion using this reduced action. The 
locally stable extrema are identified with the thermodynamic phases, while the locally unstable ones are 
identified with gravitational instantons.\\

In the rest of the section, we will carefully derive the off-shell reduced action for a five-dimensional Einstein-Maxwell theory in the 
canonical ensemble, and determine the conditions of local stability for the different saddle points. 
For the sake of simplifying the presentation, we will study the five-dimensional Euclidean Einstein-Maxwell 
theory of the Lorentzian action \eqref{eq:Action5d} with only the magnetic field turned on. 
Having an electric field does not change the physics as we discuss in \appref{app:electric charge}.

\subsection{The canonical ensemble and the saddle points}\label{sec:saddle}

Consider the Einstein-Maxwell theory with a fixed magnetic charge on a manifold $\CM$ with boundary $\Sigma$, in the canonical ensemble. 
We choose $\Sigma$ to have the topology of $S^1 \times S^1 \times S^2$, where the first $S^1$ is the thermal circle and 
the second $S^1$ is the KK circle.

In this set-up, the canonical ensemble is defined by putting the system inside a spherical cavity (i.e. a two-sphere) of radius $R_b$ and fixing the radius $\beta_t$ of the thermal circle as well as the radius $\beta_y$ of the KK circle at the boundary of the cavity. In addition, we fix the magnetic charge $Q_m$ enclosed 
in the cavity, i.e.
\be
Q_m = \frac{1}{4\pi}\,\int_{S^2_\Sigma}\, F^{(m)},
\ee
The classical action of the system is given as:
\begin{align}
I[g,F^{(m)}]=  & - \frac{1}{16\pi G_5} \, \int_{\CM} \, d^5 x\, \sqrt{g} \Big(R - \frac{1}{4} F^{(m)}_{ab} F^{(m)\,ab} \Big) - \frac{1}{8\pi G_5} \, \int_{\Sigma} d^4 x\sqrt{h}\Big(K - K_0 \Big) \nn \\
& - \frac{1}{8\pi G_5} \,\int_{\Sigma} d^4 x\sqrt{h}\, \frac{1}{2}\, n_a \, F^{(m)\,ab}\,A^{(m)}_b, \label{5dEM-EA}
\end{align}
where the last term is the standard boundary term for the Maxwell field in the canonical ensemble, with $n_a$ being the normal unit vector 
to the boundary $\Sigma$.\\

Following the general philosophy outlined above, the first step is to list the saddle points of the theory, i.e. 
solutions of Einstein's equations consistent with the boundary conditions. 
We will be concerned with the two following types of saddle points for this theory, which are the euclideanized versions of the 
black string and the topological soliton \eqref{eq:metric5d} metrics respectively.

\begin{itemize}

\item  \textbf{Euclidean Black String:} This is obtained 
by Wick rotating $t\to i \tau$ in \eqref{eq:metric5d} with $r_\text{S} > r_\text{B}$:
\begin{align}\label{BSmet-1}
& ds^2_5= (1-\frac{r_\text{S}}{r})\, d\tau^2 + (1-\frac{r_\text{B}}{r})\, dy^2 + \frac{r^2 \, dr^2}{(r- r_\text{S})(r-r_\text{B})} + r^2 d\Omega^2_2, \quad r \geq r_\text{S} > r_\text{B}, \nn \\
& F= Q_m\, \sin{\theta} \, d\theta \wedge d\phi, \qquad Q^2_m=3 r_\text{B} r_\text{S},
\end{align}
where $\tau$ is the coordinate along the thermal circle and $y$ is the coordinate along the KK circle.
For a cavity of infinite volume, regularity of the metric at the coordinate singularity $r=r_\text{S}$ fixes the asymptotic radius $\beta_t^\infty$ 
in terms of the parameters $(r_\text{B}, r_\text{S})$ as in \eqref{eq:betatinf}, 
while the corresponding radius for the KK circle $\beta^\infty_{y}$ is arbitrary. For a finite spherical cavity, the parameters $\beta_t$, 
$\beta_y$, and $Q_m$ at the boundary $\Sigma$ are given as:
\begin{equation}
 \beta_t \= \beta^\infty_t\, \sqrt{1- \frac{r_\text{S}}{R_b}},  \qquad \beta_y\= \beta^\infty_{y}\, \sqrt{1- \frac{r_\text{B}}{R_b}}, \qquad Q_m \=\sqrt{3 r_\text{B} r_\text{S}}, \label{Tol-BS1}
\end{equation}
where $r=R_b$ is the radial position of the boundary, $\beta^\infty_t$ is given in \eqref{eq:betatinf} and $\beta^\infty_{y}$ is arbitrary. 
The number of physical solutions of the black string is given by the number of real positive solutions of the equation \eqref{Tol-BS1}, 
with $r_\text{B}$ substituted in terms of $Q_m$ and $r_\text{S}$, i.e.
\be
\beta_t = \frac{2 r_\text{S}^{3/2}}{\sqrt{r_\text{S}- \frac{Q^2_m}{3r_\text{S}}}}\,\sqrt{1- \frac{r_\text{S}}{R_b}}.\label{Tol-BS3}
\ee

Note that when the boundary is set to infinity $R_b=\infty$, we retrieve the two solutions given by \eqref{eq:2BSinfiniteVol}. 
For finite $R_b$, this is a quintic equation in terms of $r_\text{S}$, while in the large $R_b$ limit it reduces to a quadratic 
equation. 

\item  \textbf{Euclidean topological soliton:} This is obtained by Wick rotating $t\to i \tau$ in \eqref{eq:metric5d} for $r_\text{B} > r_\text{S}$:
\begin{align}\label{SBmet-1}
& ds^2_5= (1-\frac{r_\text{S}}{r})\, d\tau^2 + (1-\frac{r_\text{B}}{r})\, dy^2 + \frac{r^2 \, dr^2}{(r- r_\text{S})(r-r_\text{B})} + r^2 d\Omega^2_2, \quad r \geq r_\text{B} > r_\text{S}, \nn \\
& F= Q_m\, \sin{\theta} \, d\theta \wedge d\phi, \qquad Q^2_m=3 r_\text{B} r_\text{S} . 
\end{align}
For a cavity of infinite volume, regularity of the metric at $r=r_\text{B}$ fixes the asymptotic radius $\beta^\infty_{y}$ in terms of the parameters $(r_\text{B}, r_\text{S})$ as in \eqref{eq:Ryconstraint}, while the corresponding radius for the thermal circle $\beta^\infty_t$ is arbitrary. For a finite cavity, the parameters $\beta_t$ and 
$\beta_y$ are given as:
\begin{equation}
\beta \= \beta_t^\infty\, \sqrt{1- \frac{r_\text{S}}{R_b}}, \qquad \beta_y \= \beta^\infty_{y}\, \sqrt{1- \frac{r_\text{B}}{R_b}},  \qquad Q_m \= \sqrt{3 r_\text{B} r_\text{S}}, \label{Tol-SB1}\\
\end{equation}
where $\beta^\infty_t$ is arbitrary and $\beta^\infty_{y}$ is given in \eqref{eq:Ryconstraint}. 
The number of physical solutions of the topological soliton is given by the number of real positive solutions of the equation \eqref{Tol-SB1}, 
with $r_\text{S}$ substituted in terms of $Q_m$ and $r_\text{B}$, i.e. 
\be
\beta_y = \frac{2r_\text{B}^{3/2}}{k\sqrt{r_\text{B}- \frac{Q^2_m}{3r_\text{B}}}}\,\sqrt{1- \frac{r_\text{B}}{R_b}}.
\ee

Note that this is a quintic equation in terms of $r_\text{B}$, while in the large $R_b$ limit it reduces to a quadratic 
equation. 


\end{itemize}


\subsection{Off-shell Reduced action for the Euclidean theory}
\label{sec:offShellAc}

In this section, we will compute the off-shell reduced action for the five-dimensional Euclidean Einstein-Maxwell theory in \eqref{5dEM-EA} 
subject to the appropriate boundary conditions of the canonical ensemble. In the next subsection, we will discuss 
how to extract the reduced action of the black string and the topological soliton from this answer.\\

Consider the following family of static five-dimensional metrics, with the topology $S^1 \times S^1 \times \BR_+ \times S^2$ 
in the theory \eqref{5dEM-EA} :
\be \label{metric-gen-5d-1}
ds^2 = U^2(r) dx_1^2  + V^{2}(r) dx_2^2 + \frac{1}{W^2(r)\,V^{2}(r)} dr^2 + r^2 d\Omega^2_2,
\ee
where the two circle directions are labelled by $x_i$.
In addition, we assume that the function $U(r)$ has a zero at a certain $r=r_1$, while the function 
$V(r)$ has a zero at a different $r=r_2$. We assume $ r \geq r_1 > r_2$, so that the $x_2$-circle does not 
shrink in the manifold $\cM=\{r_1\leq r\leq R_b\}$. The
$x_i$-circle has a period $2\pi \beta_i$ at the boundary $r=R_b$. Note that we are not identifying any of the circle directions as the thermal circle or the KK circle at this point.

It is convenient to write the metric and the constraints in terms of 
a new radial coordinate $\rho$, such that $\rho(r=r_1)=0$ and $\rho(r=R_b)=1$. 
The metric ansatz then takes the following form:
\be \label{metric-gen-5d-2}
ds^2 = U^2(\rho) dx_1^2 + V^{2}(\rho) dx_2^2 + \frac{1}{W^2(\rho)\,V^{2}(\rho)} d\rho^2 + r^2(\rho) d\Omega^2_2.
\ee

One can now impose the boundary conditions corresponding to canonical ensemble and the regularity condition
on the metric as follows: 

\begin{enumerate}

\item The boundary conditions of the canonical ensemble imply that :
\begin{align} \label{eq:beta12}
& \beta_1 = \frac{1}{2\pi}\,\int^{2\pi}_{0} U(1) dx_1 =  U(1), \nn \\
& \beta_{2} = \frac{1}{2\pi}\, \int^{2\pi}_{0} V(1) dx_2 =  V(1).
\end{align}

\item Regularity of the metric at the $\rho=0$ with a potential conical defect of order $k\in \mathbb{N}$:
\be \label{regcond}
[W(\rho)\, V(\rho)\, U'(\rho)]_{\rho=0} =k^{-1},
\ee
where the prime denotes derivative with respect to $\rho$.

\item The topology of the $\rho=0$ hypersurface associated with the metric \eqref{metric-gen-5d-2} 
is $S^1 \times S^1 \times S^2$. Writing the metric as
\be 
ds^2 = U^2(\rho) dx_1^2 + V^{2}(\rho) dx_2^2+ \frac{1}{(W(\rho)r'(\rho))^2\,V^{2}(\rho)} dr^2  + r^2(\rho) d\Omega^2_2,
\ee
and using the Gauss-Bonnet formula, the topology imposes a non-trivial constraint on the function $\wh{W}(\rho)$, i.e. 
it implies that the function $\wh{W}(\rho) = ({W}(\rho)\,r'(\rho))^2$ vanishes at $\rho=0$.

\end{enumerate}

The Hamiltonian and momentum constraints for the metric \eqref{metric-gen-5d-2} requires some clarification. 
From the perspective of the Euclidean action, there are two nonequivalent ways of imposing the Hamiltonian constraint. 
One may choose to foliate the five-dimensional manifold along the shrinking circle direction (i.e. along $x_1$) or along the circle direction 
which does not shrink (i.e. along $x_2$). In the first case, the Hamiltonian constraint is given by the 
$x_1 x_1$-component of the Einstein equations, while in the second case, it is given by the $x_2 x_2$-component. We will 
adopt the former choice as it yields a much simpler form of the reduced action, compared to the answer one gets 
from the second choice. However, both choices give the same reduced action in the limit where the spherical cavity is 
large, i.e. $R_b \to \infty$.

With the above choice, we note that the momentum constraints are trivially satisfied by the metric, since the off-diagonal $x_1 \mu$ components of the 
Einstein's equations vanish for the above metric ansatz. The Hamiltonian constraint -- corresponding to the 
$x_1 x_1$ component -- gives:
\be \label{HamConstr}
G^{\tau}_{\,\,\tau}  \equi R^{\tau}_{\,\,\tau}  - \frac{1}{2} R \= -\frac{1}{8}\,F^{(m)}_{\mu \nu} F^{(m)\,\,\mu \nu}.
\ee

Let us now evaluate the classical action. The individual terms in the Lagrangian can be evaluated 
for the metric ansatz \eqref{metric-gen-5d-2} as follows. First, consider the scalar curvature term: 
\begin{align}
\sqrt{g}\,\Big(R - \frac{1}{4}\,F^2\Big)= 2\,\sqrt{g}\,R^\tau_\tau= -2\sin{\theta}\,\Big( r^2 W\, U'\, V^2\Big)',
\end{align}
where in the first equality we have used the Hamiltonian constraint. 
The extrinsic curvature terms are given as:
\begin{align}
\sqrt{h}\,K= \sin{\theta}\,W\,V\,(U\,V\,r^2)', \qquad \sqrt{h}\,K_0= \sin{\theta}\,U\,V\,r^2\,(2/R_b).
\end{align}
Plugging in the above results, and using the regularity condition \eqref{regcond}, the classical action can be written in 
the simple form:
 \begin{align}
 I = \frac{2\pi^2}{G_5}\, \Big[U(1)V(1)\Big(2R_b -W(1)\,(r^2 V)'(1)\Big) - \frac{r_1^2\, V(0)}{k} \Big]. \label{RA-gen0}
 \end{align}
The equation can be simplified to the form:
 \begin{align}
 I = \frac{2\pi^2}{G_5}\, \beta_2\,\Big[\beta_1 \Big(2R_b -W(1)\,(r^2 V)'(1)\Big) -   \frac{r_1^2 V(0)}{k\,V(1)} \Big].
 \end{align}
where we have used the definitions of $\beta_1$ and $\beta_{2}$ in \eqref{eq:beta12}. Note that we have kept the functions $\{U, W, V\}$ 
completely generic (up to boundary conditions) until this point. \\

The final step is to solve the Hamiltonian constraint and evaluate the functions $W$ and $V$. Since we have a 
single constraint, we have to narrow down our ansatz further. A reasonable choice is to fix the function $V(\rho)$ which is associated 
with the non-vanishing circle direction\footnote{This condition corresponds to a choice of a slice in the phase space around the classical 
saddle point that we focus on. The resultant off-shell action will only encode variations along this restricted slice. The assumption here is that 
this choice captures all the unstable modes (if any) around the classical saddle point. One can test this assumption by 
choosing an orthogonal slice i.e. fixing $W$ instead of $V$, computing the off-shell action and checking that there are no additional unstable modes.
In fact, the two off-shell actions precisely agree in the limit of a large spherical cavity.}:
\be
V^2(\rho) = \left(1 - \frac{r_2}{r(\rho)} \right).
\ee
With this choice, the Hamiltonian constraint \eqref{HamConstr} assumes the form:
\begin{align}
(4r -3r_2)\,\wh{W}' + 4r' \wh{W} - 4r' + \frac{Q_m^2}{r^2}r'=0.
\end{align}
The above equation can be solved as:
\be
\wh{W}(\rho) = \frac{Q_m^2 + 4r^2+C r}{r(4r-3 r_2)},
\ee
where $C$ is an integration constant. The boundary condition $\wh{W}(0) =0$ fixes the constant $C$, and the solution becomes:
\be
\wh{W}(\rho) = \Big(1 - \frac{r_1}{r(\rho)}  \Big)\,\frac{4 r_1 r(\rho) -Q_m^2}{4 r(\rho)-3r_2}.
\ee

Plugging in $V$ and $W$, we have the following classical action : 
\begin{align}
& I \= \frac{2\pi^2}{G_5}\,\beta_2\, \Big[\beta_1 \, A(r_1, r_2, Q_m, R_b) - B(r_1, r_2)\Big] , \label{RA-gen2} \\
& A(r_1, r_2, Q_m, R_b) = 2 R_b- \frac{1}{2}\,\sqrt{ \frac{\left(R_b-r_1 \right)(4 R_b -3 r_2)\left(4 r_1 R_b -Q_m^2 \right)}{r_1(R_b-r_2)}} , \\
& B(r_1, r_2) = \frac{r_1^2}{k}\, \sqrt{\frac{R_b(r_1- r_2)}{r_1(R_b-r_2)} }.
\end{align}

The reduced action is then a function of two variables, $(r_1,r_2)$, with the parameters $\beta_1$, $\beta_2$, $R_b$, 
and $Q_m$ being held fixed. 

\subsection{The extrema and local stability}

The locally-stable Euclidean saddle points are the local minima of the off-shell reduced action. 
First, the variables $(r_1,r_2)$ at the extrema of the action are given in terms of the boundary data as follows:
\begin{equation}
\partial_{r_1} I \= \partial_{r_2} I \= 0 \qquad \Rightarrow \qquad Q_m^2 \= 3 r_1 r_2\,,\qquad  \beta_1 = \frac{2 r^{3/2}_1}{k\sqrt{r_1- r_2}}\sqrt{1-\frac{r_1}{R_b}}.
\label{eq:extremaCond}
\end{equation}
The conditions of the extrema match the ones obtained for the two topological soliton and black string saddle points highlighted in the previous section. However, we remind the reader that at this point there are no distinctions between the $x_1$ and $x_2$ circles, and the extrema obtained from the conditions above do not correspond to these solutions yet. 

The stability of the saddle points requires that the Hessian of the potential at the extrema is positive definite, where the Hessian is defined by $\partial_{r_a} \partial_{r_b}  I$. For a $2 \times 2$ Hessian, this is guaranteed if one element of the diagonal and the determinant is positive. 
Therefore, the conditions for local stability at saddle points are given by
\begin{equation}
\begin{split}
&\partial_{r_2}^2  I \bigl|_\text{extrema} \= \frac{\pi^2 \beta_2}{2 G_5} \,\sqrt{\frac{R_b r_1^3}{(r_1-r_2)^3(R_b-r_2)^3}} \,\frac{(4 R_b-3 r_1)(R_b-r_2)}{(4 R_b - 3 r_2)} >0\,,\\
&\partial_{r_1}^2  I \,\partial_{r_2}^2  I - (\partial_{r_1}\partial_{r_2} I)^2 \bigl|_\text{extrema} \= \left( \frac{2\pi^2 \beta_2}{ G_5} \right)^2 \frac{2 R_b(R_b-r_1)r_1}{(4 R_b-3 r_2)(R_b-r_2)^2(r_1-r_2)^2}\\
&\hspace{6.5cm} \times \left( R_b(4 r_2-2 r_1)+r_1(3 r_1-5 r_2)\right)>0,
\end{split}
\label{eq:HessianCond}
\end{equation}
where $(r_1,r_2)$ are functions of $(Q_m, \beta_1)$ given by \eqref{eq:extremaCond}. It is clear that in the regime we are in, $R_b > r_1 > r_2>0$, $\partial_{r_2}^2  I \bigl|_\text{extrema}>0$ is trivially satisfied. The condition for having a locally stable saddle point is therefore given by
\begin{equation}
 R_b(4 r_2-2 r_1)+r_1(3 r_1-5 r_2)>0\,, \qquad \implies \quad 3r_1^2 - 5r_1r_2 - 2r_1 R_b + 4R_br_2 >0.
\end{equation}
Eliminating the variable $r_2$ in terms of the magnetic charge, the above equation can be written as
\be
9r_1^3 - 6r^2_1R_b - 5r_1Q_m^2  + 4R_b Q_m^2  >0. \label{eq:StabCond0}
\ee

In the limit of a large box, i.e. $R_b \gg r_1 > r_2$, the above equation simplifies to
\begin{align} 
  2 Q_m^2  - 3r^2_1 >0, 
\end{align}
and the constraint on $r_1$ can be written in the two following equivalent ways:
\be \label{eq:StabCond}
\sqrt{\frac{1}{3}}\,Q_m < r_1 < \sqrt{\frac{2}{3}}\,Q_m \quad \Longleftrightarrow \quad r_2 < r_1 < 2r_2.
\ee

Therefore, we have to study all the solutions obtained from \eqref{eq:extremaCond} for different regimes of $(Q_m, \beta_1)$ and discuss which ones satisfy \eqref{eq:StabCond}. Before doing so, we first derive the reduced action/free energy for the black string and the topological soliton by using the generic expression \eqref{RA-gen2} and analytically continuing the Euclidean metric \eqref{metric-gen-5d-1} to an Lorentzian space-time in an appropriate fashion.

\subsection{Free energy for the black string and topological soliton}

In this section, we derive the free energy for the black string and the effective potential for the topological soliton 
from the reduced action \eqref{RA-gen2}. This is done by taking two different analytic continuations of the 
five-dimensional Euclidean metric in \eqref{metric-gen-5d-1}.

%

\subsubsection*{Black String} 

We consider the analytic continuation of the metric \eqref{metric-gen-5d-1} where 
$x_1 \to it$, and $x_2 \to y$, and the absence of conical defect at $r=r_1$, $k=1$. In this case, the metric in the Lorentzian space-time is a black string \eqref{eq:metric5d} with \eqref{eq:BScond}, 
and therefore one should identify the $x_1$-direction as the thermal circle and the $x_2$-direction 
as the KK circle. We then have the following identification:
\begin{align}
& \beta_{1} = \beta_t, \qquad \beta_2 = \beta_{y}, \qquad k=1, \nonumber\\
& r_1 = r_\text{S}, \qquad r_2 = r_\text{B}, \quad \text{with} \quad r_\text{S} > r_\text{B}. \label{eq:ContinuationBS}
\end{align}
With this identification, the off-shell action \eqref{RA-gen2} is:
\begin{align}
I_\text{BS} = & \frac{2\pi^2}{G_5}\,\beta_y\, \Biggl[\beta_t \left(2 R_b- \frac{1}{2}\,\sqrt{ \frac{\left(R_b-r_\text{S} \right)(4 R_b -3 r_\text{B})\left(4 r_\text{S} R_b -Q_m^2 \right)}{r_\text{S}(R_b-r_\text{B})}} \right) 
- r_\text{S}^2\, \sqrt{\frac{R_b(r_\text{S}- r_\text{B})}{r_\text{S}(R_b-r_\text{B})} }\Biggr] .\label{eq:EP-BS}
\end{align}

In this case, the stability condition \eqref{eq:StabCond} translates to :
\be \label{eq:StabCondBS}
r_\text{B} < r_\text{S} < 2r_\text{B}.
\ee

Note that the above condition for quantum mechanical stability is precisely the same as the condition for classical stability,
as mentioned in \Secref{sec:review0} and derived in detail in \appref{app:GL-TS}. 

\subsubsection*{Topological Soliton}

Consider the analytic continuation of the metric \eqref{metric-gen-5d-1} where 
$x_1 \to y$, and $x_2 \to it$. In this case, the metric in the Lorentzian space-time is the topological soliton \eqref{eq:metric5d} with \eqref{eq:TScond},
and therefore one should identify the $x_2$-direction as the thermal circle and the $x_1$-direction 
as the KK circle. We then have the following identification:
\begin{align}
& \beta_{1} = \beta_{y}, \qquad \beta_2 = \beta_t, \nonumber \\
& r_1 = r_\text{B}, \qquad r_2 = r_\text{S}, \quad \text{with} \quad r_\text{B} > r_\text{S}. \label{eq:ContinuationTS}
\end{align}
With this identification, the off-shell action is:
\begin{align}
I_\text{TS} = & \frac{2\pi^2}{G_5}\,\beta_t\, \Biggl[\beta_y \left(2 R_b- \frac{1}{2}\,\sqrt{ \frac{\left(R_b-r_\text{B} \right)(4 R_b -3 r_\text{S})\left(4 r_\text{B} R_b -Q_m^2 \right)}{r_\text{B}(R_b-r_\text{S})}} \right) 
-\frac{r_\text{B}^2}{k}\, \sqrt{\frac{R_b(r_\text{B}- r_\text{S})}{r_\text{B}(R_b-r_\text{S})} }\Biggr] .\label{eq:EP-TS}
\end{align}

In this case, the stability condition \eqref{eq:StabCond} translates to :
\be \label{eq:StabCondTS}
r_\text{S} < r_\text{B} < 2r_\text{S}.
\ee

This implies that in the regime $r_\text{B} > 2r_\text{S}$, the topological soliton has a quantum mechanical instability, 
in addition to the classical instability which was discussed in \Secref{sec:CS}. \\

The two off-shell actions are therefore formally related as $I_\text{BS} \leftrightarrow I_\text{TS}$,
when the parameters $\beta_t \leftrightarrow k\,\beta_{y}$ and $r_\text{B}  \leftrightarrow r_\text{S}$. 
Therefore, the thermodynamic phases and their stability condition for the bubbles 
will work out in a way similar to the black string, with $k\,\beta_{y}$ playing the role of $\beta_t$.\\

\section{Thermodynamic phases in the infinite box limit}\label{sec:infbox}
In this section, we will study the thermodynamics of the five-dimensional Einstein-Maxwell theory in the canonical ensemble, where the 
size of the box is infinitely large, i.e. $R_b \to \infty$. The infinite box limit is convenient as it drastically simplifies the analysis 
and still contains most of the physics. We will first discuss the thermodynamic phases for the black string and the topological solitons
and their local stability. We will then address the question of global stability and discuss the Hawking-Page transitions between 
the different locally stable phases.

In the infinite box limit, the reduced actions describing the black string phase and the topological soliton phases are now given by 
\begin{equation}
\begin{split}
I_\text{BS}^\infty &\= \frac{2\pi^2\,\beta_y^\infty}{G_5} \left[\beta_t^\infty \,\frac{4 r_\text{S}^2 + Q_m^2 - r_\text{S} r_\text{B}}{4r_\text{S}} \- r_\text{S}^\frac{3}{2} \sqrt{r_\text{S}-r_\text{B}}\right]\,, \\
I_\text{TS}^\infty &\= \frac{2\pi^2\,\beta_t^\infty}{G_5} \left[\beta_y^\infty \,\frac{4 r_\text{B}^2 + Q_m^2 - r_\text{S} r_\text{B}}{4r_\text{B}} \- \frac{r_\text{B}^\frac{3}{2}}{k} \sqrt{r_\text{B}-r_\text{S}}\right]\,, 
\label{eq:EffPotInf}
\end{split}
\end{equation}
where we have introduced ``$\infty$'' index to highlight that we are in infinite box limit.

\subsection{Stability of the black string phases}

\begin{figure}[h]\centering
\begin{subfigure}{0.47\textwidth}\centering
 \includegraphics[width=\textwidth]{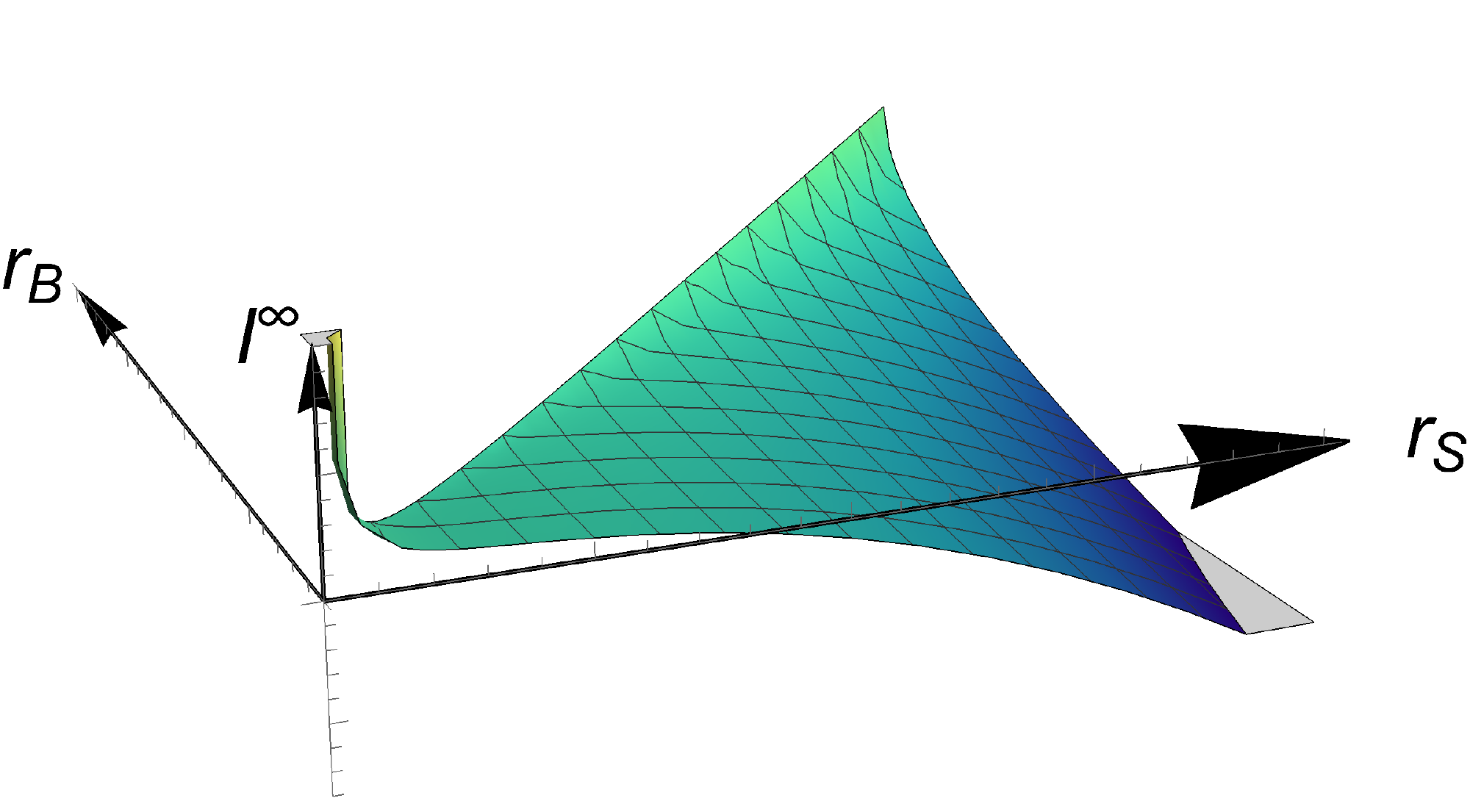}
 \caption{$4 Q_m > \sqrt{3} \beta_t^\infty$}
 \end{subfigure}
 \hspace{0.2cm}
 \begin{subfigure}{0.47\textwidth}\centering
   \includegraphics[width=\textwidth]{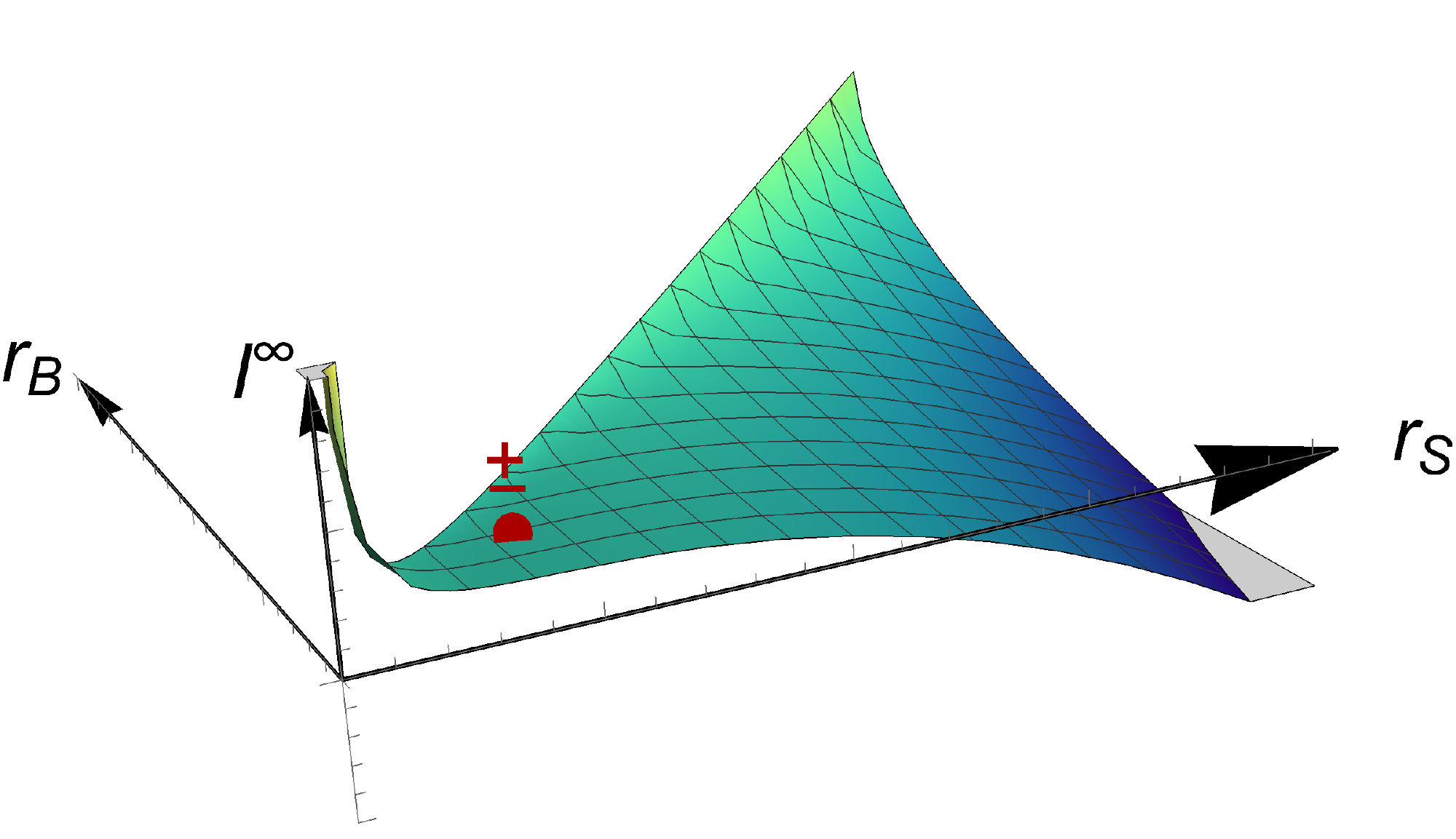}
    \caption{$4 Q_m = \sqrt{3} \beta_t^\infty$}
 \end{subfigure}
 \begin{subfigure}{0.47\textwidth}\centering
 \includegraphics[width=\textwidth]{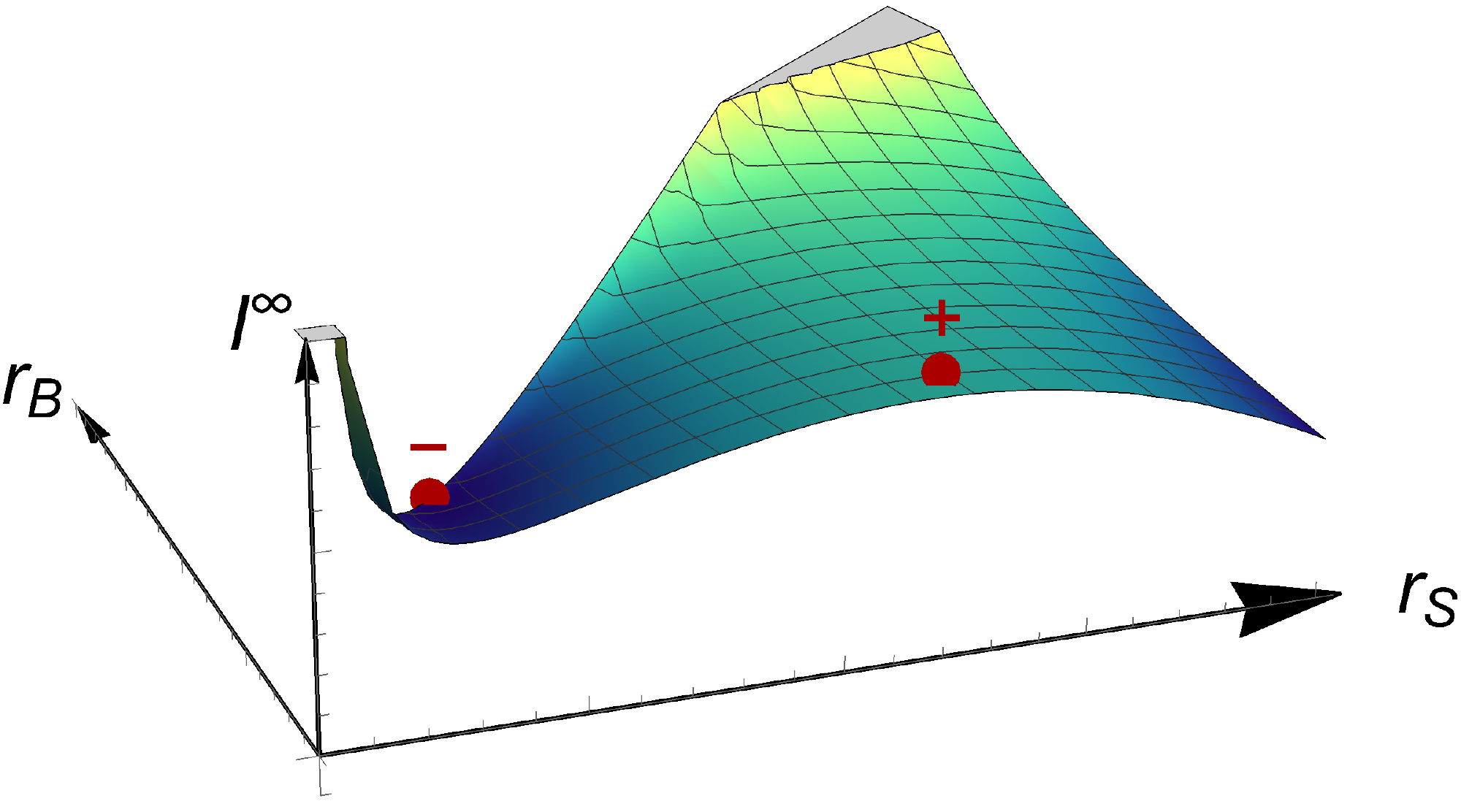}
 \caption{$4 Q_m < \sqrt{3} \beta_t^\infty$}
 \end{subfigure}
 \caption{Plots of $I^\infty_\text{BS}$ in terms of $(r_\text{S},r_\text{B})$ for three regimes of $(Q_m,\beta_t^\infty)$: when there are no extrema (a), one extremum (b) and two extrema (c).}
 \label{fig:BSpot}
\end{figure}

The extrema of the black string reduced action can be read off from \eqref{eq:extremaCond} given the identification \eqref{eq:ContinuationBS}. 
These extrema manifestly coincide with the classical solutions already studied in section \ref{sec:BSLorentz}, and can be classified as follows 
for different regimes of $(Q_m, \beta_t^\infty )$:

\begin{itemize}

\item For $4 Q_m > \sqrt{3} \beta_t^\infty$, there is no extremum, as shown in Fig.\ref{fig:BSpot}(a). This implies that no black string solution 
exists in this regime of parameters $(Q_m, \beta_t^\infty )$.

\item For $4 Q_m < \sqrt{3} \beta_t^\infty$, we have two extrema, given by  
$(r_{\text{S}\,+},r_{\text{B}\,+})$ and $(r_{\text{S}\,-},r_{\text{B}\,-})$ respectively,
as shown in Fig.\ref{fig:BSpot}(c).
\begin{equation}
r_{\text{S} \,\pm} \= \frac{\beta^\infty_t}{2\sqrt{2}} \sqrt{1 \pm \sqrt{1-\frac{16 Q_m^2}{3{\beta^\infty_t}^2}}}\,,\qquad r_{\text{B} \,\pm} \= \frac{Q_m}{\sqrt{6}} \sqrt{1 \mp \sqrt{1-\frac{16 Q_m^2}{3{\beta^\infty_t}^2}}}\,.
\label{eq:2BSinfiniteBox}
\end{equation}

These correspond to two black string solutions of different sizes, where the larger one is denoted as ``$+$'' and the 
smaller one is denoted as ``$-$''. From \figref{fig:BSpot}(c), it is clear that only the small black string corresponds to a local minimum. 
One can check that the stability condition \eqref{eq:StabCondBS} is always satisfied for the smaller black string parametrized by 
$(r_{\text{S}\,-},r_{\text{B}\,-})$, and violated for the larger black string $(r_{\text{S}\,+},r_{\text{B}\,+})$. 
However, because the potential goes to $-\infty$ as $r_\text{S}$ gets large, the smaller black string is only a local minimum 
of the effective potential and therefore corresponds to a meta-stable phase. The larger black string acts as a gravitational 
potential mediating this process.

\item For $4 Q_m = \sqrt{3} \beta_t^\infty$, $r_{\text{S}\,\pm}=2r_{\text{B}\,\pm} = \frac{\beta^\infty_t}{2\sqrt{2}}$ and the Hessian degenerates at the extremum \eqref{eq:HessianCond}. Therefore, this extremum is neither a maximum nor a minimum, and does not qualify as a thermodynamic phase.

\end{itemize}

Note that the result is consistent with the known results for the neutral limit $Q_m=0$, which correspond to pure five-dimensional Einstein gravity in a canonical ensemble \cite{Gross:1982cv}. 
In this case, the small black string reduces to the hot KK space-time as $r_{\text{S}\,-}=r_{\text{B}\,-}=0$, and the large black string reduces to the S$^1$ fibration over an Euclidean Schwarzschild black hole. As is well known, the latter acts as a gravitational instanton in the theory 
and mediates the nucleation of black strings from the hot KK space-time, thereby rendering it unstable \cite{Gross:1982cv}.

\begin{figure}[h]
\centering
\includegraphics[width=0.95\textwidth]{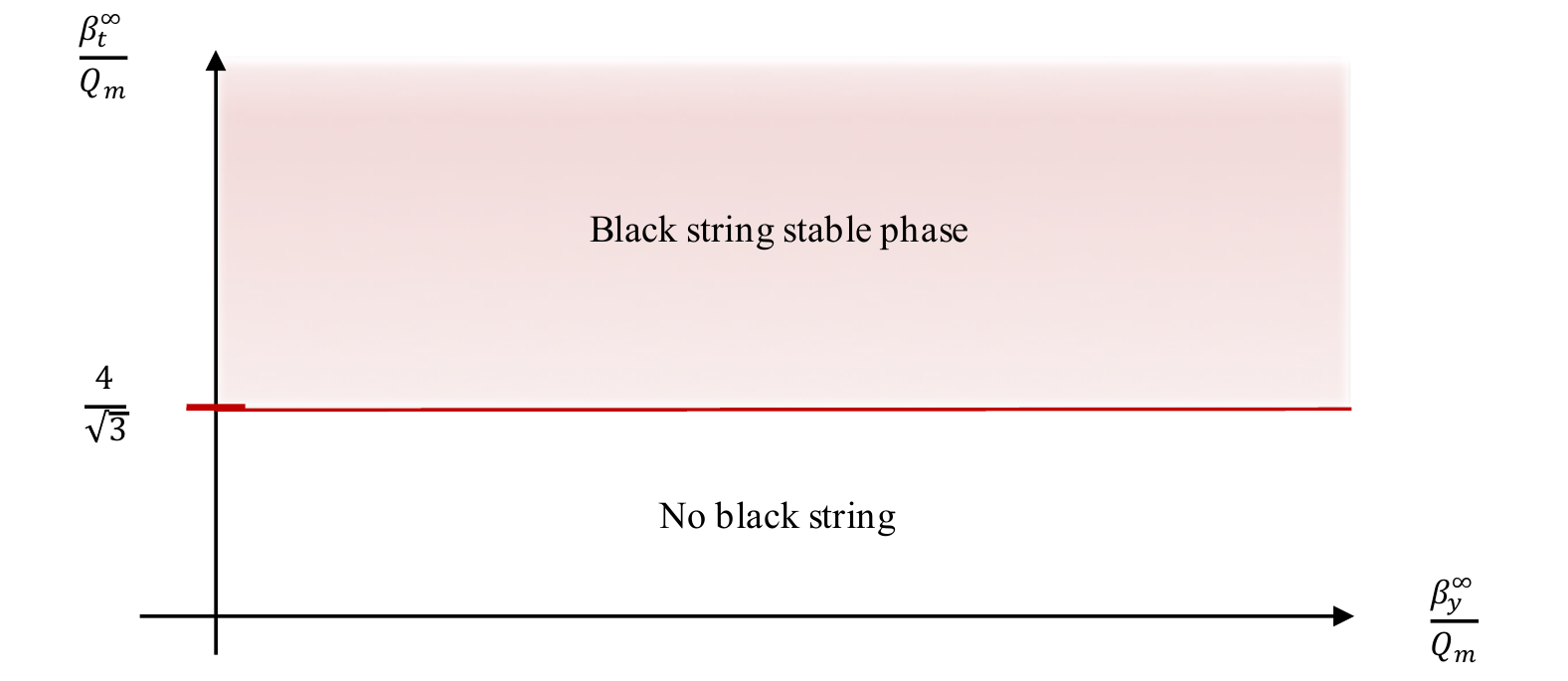}
\caption{Phase space of locally-stable magnetic black string.}
\label{fig:phaseBS}
\end{figure}

In \figref{fig:phaseBS}, we summarize the phase space of locally-stable magnetic black string in the $(\beta_t^\infty,\beta_y^\infty)$ plane in units of $Q_m$. 

The reduced actions at the two extrema for $4 Q_m < \sqrt{3} \beta_t^\infty$ are given as:
\begin{align}
I^\infty_\text{BS} \bigl|_{\pm \text{extremum}} \= \frac{\pi^2 \beta_y^\infty {\beta_t^\infty}^2}{2\sqrt{2} G_5} \left(2 \mp \sqrt{1-\frac{16 Q_m^2}{3{\beta_t^\infty}^2}} \right) \,\sqrt{1\pm \sqrt{1-\frac{16 Q_m^2}{3{\beta_t^\infty}^2}}}. \label{BS-2ext}
\end{align}

The behavior of the reduced action at large $\beta_t^\infty$ is given as :
\begin{align}
& I^\infty_\text{BS} \bigl|_{+ \text{extremum}} \xrightarrow[]{\frac{\beta_t^\infty}{Q_m} \gg 1}  \frac{\pi^2 \beta_y^\infty {\beta_t^\infty}^2}{2 G_5} 
+O\Big((\frac{Q_m}{\beta_t^\infty})^0\Big),\\
& I^\infty_\text{BS} \bigl|_{- \text{extremum}} \xrightarrow[]{\frac{\beta_t^\infty}{Q_m} \gg 1}  \frac{\sqrt{3}\pi^2 \beta_y^\infty {\beta_t^\infty} Q_m}{G_5} 
+O\Big((\frac{Q_m}{\beta_t^\infty})\Big).
\end{align}

From the above expansions, one observes that the free energy of the smaller black string is constant at large $\beta_t^\infty$ 
(for a fixed $Q_m$), while for the larger black string it grows linearly.\\

As mentioned earlier, the larger black string mediates a decay of the meta-stable smaller black string. The nucleated black string 
is thermodynamically unstable and can either Hawking-evaporate to get back to the smaller black string, or can grow in size
\footnote{It is not clear from our semi-classical analysis what the end-point of this growing black string should be.}. 
The rate of nucleation is given as:
\be
\Gamma = \Gamma_0\, e^{- \Big( I^\infty_\text{BS} \bigl|_{+ \text{extremum}} - I^\infty_\text{BS} \bigl|_{- \text{extremum}} \Big)},
\ee
where $ I^\infty_\text{BS} \bigl|_{\pm \text{extremum}}$ are given in \eqref{BS-2ext}. 
The function $( I^\infty_\text{BS} \bigl|_{+ \text{extremum}} - I^\infty_\text{BS} \bigl|_{- \text{extremum}})$ is a 
positive definite monotonic increasing function of $\beta_t^\infty$ (and $\frac{\beta_t^\infty}{Q_m}$).

At large $\frac{\beta_t^\infty}{Q_m}$, one expects the smaller string to become more and more stable. In this case, the nucleation rate is exponentially suppressed such as
\be
\Gamma = \Gamma_0\, e^{- \frac{\pi^2 \beta_y^\infty {\beta_t^\infty} Q_m}{2G_5}\Big(\frac{\beta_t^\infty}{Q_m} - 2\sqrt{3}\Big)} 
\approx \Gamma_0\, e^{- \frac{\pi^2 \beta_y^\infty {\beta_t^\infty}^2}{2G_5}}. \label{eq:rateBS}
\ee
The temperature dependence of the exponent in this case, is exactly the same as that for a four-dimensional 
Schwarzschild black hole, found by Gross-Perry-Yaffe \cite{Gross:1982cv}. Indeed,  taking large $\frac{\beta_t^\infty}{Q_m}$ is equivalent to considering the neutral limit where the black string becomes a trivial S$^1$ fibration over a Schwarzschild black hole.

As the parameter $\frac{\beta_t^\infty}{Q_m}$ approaches the minimal value $\frac{4}{\sqrt{3}}$, the smaller string is expected to become 
increasingly unstable. This can be seen from the fact that in the limit $\frac{\beta_t^\infty}{Q_m} \to \frac{4}{\sqrt{3}}$, the nucleation rate is no longer exponentially suppressed and is given by $\Gamma_0$. Therefore, the meta-stable black string  suffers from a quantum-mechanical instability with a high probability of nucleating an unstable black string. At the same time, the smaller string approaches the classical instability threshold, $2r_\text{B}\gtrsim r_\text{S}$, as detailed in section \ref{sec:CS}. It is expected to suffer increasingly from long-lived classical perturbations that should start looking like Gregory-Laflamme modes.

\subsection{Stability of the topological soliton phases}

The thermodynamic phases of the topological solitons and their local stability can be analyzed in a fashion similar to the black string.
As discussed in section \ref{sec:BSLorentz}, there exists a tower of pairs of extrema labeled by the positive integer $k$ 
for any boundary data $(Q_m,\beta_y^\infty,\beta_t^\infty)$, where the parameters $(r^{(k)}_{\text{B} \,\pm}, r^{(k)}_{\text{S} \,\pm})$ are given by:
\begin{equation}
r^{(k)}_{\text{B} \,\pm} \= \frac{k\,\beta^\infty_y}{2\sqrt{2}\,} \sqrt{1 \pm \sqrt{1-\frac{16 Q_m^2}{3k^2\,{\beta^\infty_y}^2}}}\,,\qquad r^{(k)}_{\text{S} \,\pm} \= \frac{Q_m}{\sqrt{6}} \sqrt{1 \mp \sqrt{1-\frac{16 Q_m^2}{3k^2\,{\beta^\infty_y}^2}}}\,.
\end{equation}
The tower starts at $k\geq k_\text{min}=\lfloor \frac{4 Q_m}{\sqrt{3} \beta_y^\infty} \rfloor +1$ where $\lfloor x \rfloor$ is the integer part of $x$. 

For a given $k$, the behavior of the potential is similar to \figref{fig:BSpot} if we relabel the axes, $r_\text{S} \leftrightarrow r_\text{B}$, and 
take $\beta_t^\infty \leftrightarrow k \beta_y^\infty$. The reduced actions in \eqref{eq:EffPotInf} transform as: 
\be \label{BS-TSmap}
I^\infty_\text{BS} \leftrightarrow  I^\infty_\text{TS}, \qquad \text{for} \quad  (\beta_t^\infty,r_\text{S}) \leftrightarrow (k\beta_y^\infty,r_\text{B}).
\ee 

Therefore, in the regime $4 Q_m > \sqrt{3} k \beta_y^\infty$, there exists no topological soliton solution for the given label $k$. 
In the  regime $4 Q_m < \sqrt{3} k \beta_y^\infty$, the smaller topological soliton, parametrized by $(r^{(k)}_{\text{S} \,-}, r^{(k)}_{\text{B} \,-} )$, 
is the locally-stable phase while the larger one, parametrized by $(r^{(k)}_{\text{S} \,+}, r^{(k)}_{\text{B} \,+} )$, corresponds to a gravitational instanton.
Finally, for $4 Q_m = \sqrt{3} k \beta_y^\infty$, the two extrema merge to give a saddle point.
  
\begin{figure}[h]
\centering
\includegraphics[width=0.9\textwidth]{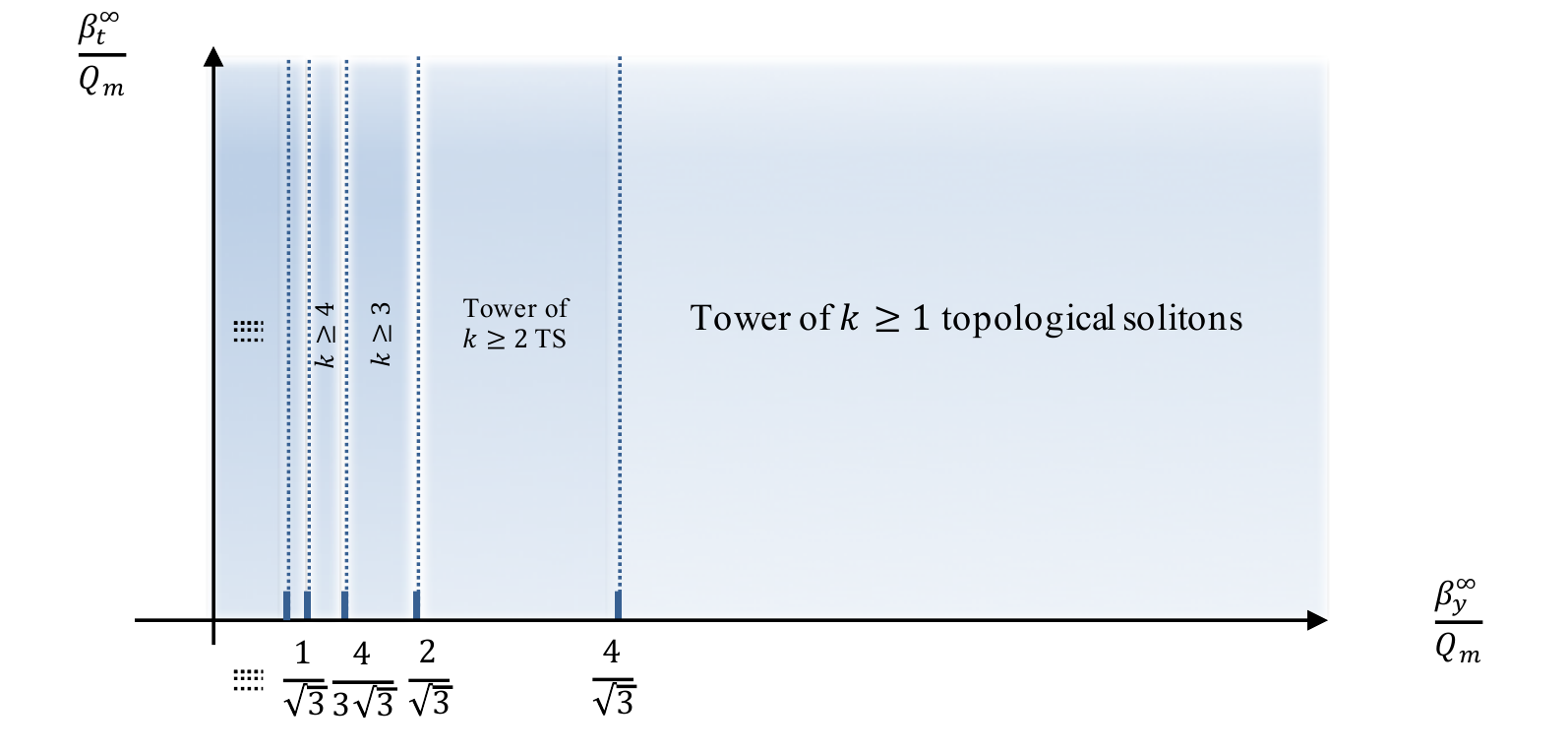}
\caption{Phase space of locally-stable topological solitons.}
\label{fig:phaseTS}
\end{figure}

In \figref{fig:phaseTS}, we summarize the phase space of locally-stable topological soliton in the $(\beta_t^\infty,\beta_y^\infty)$ plane in units of 
$Q_m$. Note that the phase space of topological solitons is qualitatively different from that of the black string, given in \figref{fig:phaseBS}.
In the latter case, there is a lower bound of the parameter $\frac{\beta_t^\infty}{Q_m}$ for a black string phase to exist. In the former case, 
there always exists a tower of topological soliton solutions for any value of the parameter $\frac{\beta_y^\infty}{Q_m}$. The tower starts at 
$k_\text{min}=1$ if $\frac{\beta_y^\infty}{Q_m} > \frac{4}{\sqrt{3}}$, and at $k_\text{min}>1$ if $\frac{\beta_y^\infty}{Q_m} \leq \frac{4}{\sqrt{3}}$, 
as shown in \figref{fig:phaseTS}.

In the limit of vanishing magnetic charge, the smaller topological soliton degenerates to the hot KK space-time while the larger one 
becomes a static bubble with a conical defects. The latter is a gravitational instanton that mediates the decay of the KK spacetime.
One can think of it as the gravitational instanton analogous to Witten's bubble-of-nothing \cite{Witten:1981gj}, for the case of a 
hot KK space.
 
 
The reduced action at the different extrema for a given $k$ are :
\begin{align} 
 I^\infty_\text{TS} \bigl|_{k\pm \text{extremum}} \= \frac{k\pi^2 \beta_t^\infty {\beta_y^\infty}^2}{2\sqrt{2} G_5} \left(2 \mp \sqrt{1-\frac{16 Q_m^2}{3k^2{\beta_y^\infty}^2}} \right) \,\sqrt{1\pm \sqrt{1-\frac{16 Q_m^2}{3k^2{\beta_y^\infty}^2}}}. \label{TS-2ext}\\
\end{align}

In the regime $4 Q_m < \sqrt{3} k \beta_y^\infty$, the reduced actions at large $\beta_y^\infty$ are given as:
\begin{align}
& I^\infty_\text{TS} \bigl|_{k+ \text{extremum}} \xrightarrow[]{\frac{k\beta_y^\infty}{Q_m} \to \infty}  \frac{k\pi^2 \beta_t^\infty {\beta_y^\infty}^2}{2 G_5} 
+O\Big((\frac{Q_m}{k\beta_y^\infty})^0\Big),\\
& I^\infty_\text{TS} \bigl|_{k- \text{extremum}} \xrightarrow[]{\frac{k\beta_y^\infty}{Q_m} \to \infty}  \frac{\sqrt{3}\pi^2 \beta_y^\infty {\beta_t^\infty} Q_m}{G_5} 
+O\Big((\frac{Q_m}{k\beta_y^\infty})\Big).
\end{align}

Analogous to what we found for the black string, the larger topological soliton can mediate a decay 
of the smaller topological soliton. The nucleated topological soliton is unstable and can either contract to get back to the 
smaller soliton, or can grow in size\footnote{It is not clear from our semi-classical analysis what the end-point of this 
growing topological soliton should be.}. The rate of nucleation is given as:
\be
\Gamma = \Gamma_0\, e^{- \Big( I^\infty_\text{TS} \bigl|_{+ \text{extremum}} - I^\infty_\text{TS} \bigl|_{- \text{extremum}} \Big)},
\ee
where $ I^\infty_\text{TS} \bigl|_{\pm \text{extremum}}$ are given in \eqref{TS-2ext}. The function 
$( I^\infty_\text{TS} \bigl|_{+ \text{extremum}} - I^\infty_\text{TS} \bigl|_{- \text{extremum}} \Big)$ is a positive 
definite monotonic increasing function $k\beta_y^\infty$ (and $\frac{k\beta_y^\infty}{Q_m}$).

At large $\frac{k\beta_y^\infty}{Q_m}$, one expects the smaller soliton to become more and more stable. In this case, the nucleation rate is exponentially suppressed and given by
\be
\Gamma = \Gamma_0\, e^{- \frac{\pi^2 \beta_t^\infty {\beta_y^\infty} Q_m}{2G_5}\Big(\frac{k\beta_y^\infty}{Q_m} - 2\sqrt{3}\Big)} 
\approx \Gamma_0\, e^{- \frac{k\pi^2 \beta_t^\infty {\beta_y^\infty}^2}{2G_5}}. \label{eq:rateTS}
\ee

As the parameter $\frac{k\beta_y^\infty}{Q_m}$ approaches the minimal value $\frac{4}{\sqrt{3}}$, the smaller soliton is expected to become 
increasingly unstable. This can be seen from the fact that in the limit $\frac{k\beta_y^\infty}{Q_m}\to \frac{4}{\sqrt{3}}$, the nucleation rate is no longer exponentially suppressed and is given by $\Gamma_0$. Therefore, the meta-stable soliton suffers from a quantum-mechanical instability with a high probability of nucleating the unstable soliton. At the same time, the smaller soliton approaches the classical instability threshold, $2r_\text{S}\gtrsim r_\text{B}$, as detailed in section \ref{sec:CS}. It is expected to suffer increasingly from long-lived classical perturbations that should start looking like Gregory-Laflamme modes.

\subsection{Hawking-Page phase transitions}
\label{sec:HPinfinitebox}

As explained in the beginning of \Secref{sec:review}, a locally-stable phase can undergo a Hawking-Page transition to another 
locally-stable phase, if the latter has a smaller free energy. Generically, for a given set of boundary data, we have a tower of 
locally-stable topological soliton phases on the one hand, and a single locally-stable black string phase on the other. 
Therefore, we have two possible transitions in the system: a topological soliton/ topological soliton transition 
and a topological soliton/ black string transition, which we will now analyze to obtain the globally-stable phase 
diagram in the $(\frac{\beta_y^\infty}{Q_m}, \frac{\beta_t^\infty}{Q_m})$-plane.

\begin{itemize}
\item[•] \underline{topological soliton $\leftrightarrow$ topological soliton:}

We consider two locally-stable topological solitons, with orbifold parameter $k_1$ and $k_2$ respectively, 
for fixed boundary data $(Q_m,\beta_t^\infty,\beta_y^\infty)$. Their free energy, obtained from \eqref{TS-2ext}, is
\begin{equation} \label{FE-stabTS}
F_{\text{TS}\,a} \= \frac{k_a \pi^2 {\beta_y^\infty}^2}{2\sqrt{2} G_5} \left(2 + \sqrt{1-\frac{16 Q_m^2}{3k_a^2{\beta_y^\infty}^2}} \right) \,\sqrt{1- \sqrt{1-\frac{16 Q_m^2}{3k_a^2{\beta_y^\infty}^2}}},\qquad a=1,2.
\end{equation}
One observes that $F_{\text{TS}\,1} > F_{\text{TS}\,2}$, if and only if $k_1 > k_2$. This implies that the topological soliton with the smallest 
possible conical defect , $ k=k_\text{min}=\lfloor \frac{4 Q_m}{\sqrt{3} \beta_y^\infty} \rfloor +1$, is the globally-stable topological soliton 
phase. Therefore, any locally-stable topological soliton with $k > k_\text{min}$ will undergo a Hawking-Page transition to the $k=k_\text{min}$ soliton.

\item[•] \underline{topological soliton $\leftrightarrow$ black string:}

We consider now the topological soliton with conical defect $k=k_\text{min}$ and a locally-stable black string, for the fixed 
boundary data $(Q_m,\beta_t^\infty,\beta_y^\infty)$. The free energy of the topological soliton is given by \eqref{FE-stabTS} 
with $k=k_\text{min}$, while that for the black string is
\begin{equation}
F_{\text{BS}} \= \frac{\pi^2 \beta_y^\infty {\beta_t^\infty}}{2\sqrt{2} G_5} \left(2 + \sqrt{1-\frac{16 Q_m^2}{3{\beta_t^\infty}^2}} \right) \,\sqrt{1- \sqrt{1-\frac{16 Q_m^2}{3{\beta_t^\infty}^2}}}. 
\end{equation}
The two functions $F_{\text{TS}}$ and $F_{\text{BS}}$ are of the following form:
\begin{align}
& {\beta_y^\infty}^{-1}F_{\text{TS}} = f(u,Q_m) , \qquad {\beta_y^\infty}^{-1}F_{\text{BS}} = f(v,Q_m),\\
& f(x,Q_m)= \frac{\pi^2 x}{2\sqrt{2} G_5} \left(2 + \sqrt{1-\frac{16 Q_m^2}{3x^2}} \right) \, \sqrt{1- \sqrt{1-\frac{16 Q_m^2}{3x^2}}},\\
& u= k_\text{min}{\beta_y^\infty}, \qquad v= {\beta_t^\infty}.
\end{align}
Noting that $f(x,Q_m)$ is a positive definite monotonic increasing function, we conclude that 
\begin{equation}
F_\text{BS} \lesseqgtr F_\text{TS} \,,\quad \text{if and only if} \quad \beta_t^\infty  \lesseqgtr k_\text{min} \beta_y^\infty .
\end{equation} 
Therefore, we conclude that the topological soliton with the smallest conical defect is the globally-stable phase 
in the low-temperature regime of boundary data, i.e.
\be
\beta_t^\infty > k_\text{min} \beta_y^\infty >\frac{4}{\sqrt{3}}Q_m,
\ee
and the locally-stable black string will undergo a Hawking-Page transition to the topological soliton.
On the other hand, the black string is the globally-stable phase in the high-temperature regime of boundary data, i.e.
\be
k_\text{min} \beta_y^\infty > \beta_t^\infty >\frac{4}{\sqrt{3}}Q_m,
\ee
and the topological soliton will undergo a Hawking-Page transition to the black string in this regime. 

\end{itemize}

\begin{figure}[h]
\centering
\includegraphics[width=0.9\textwidth]{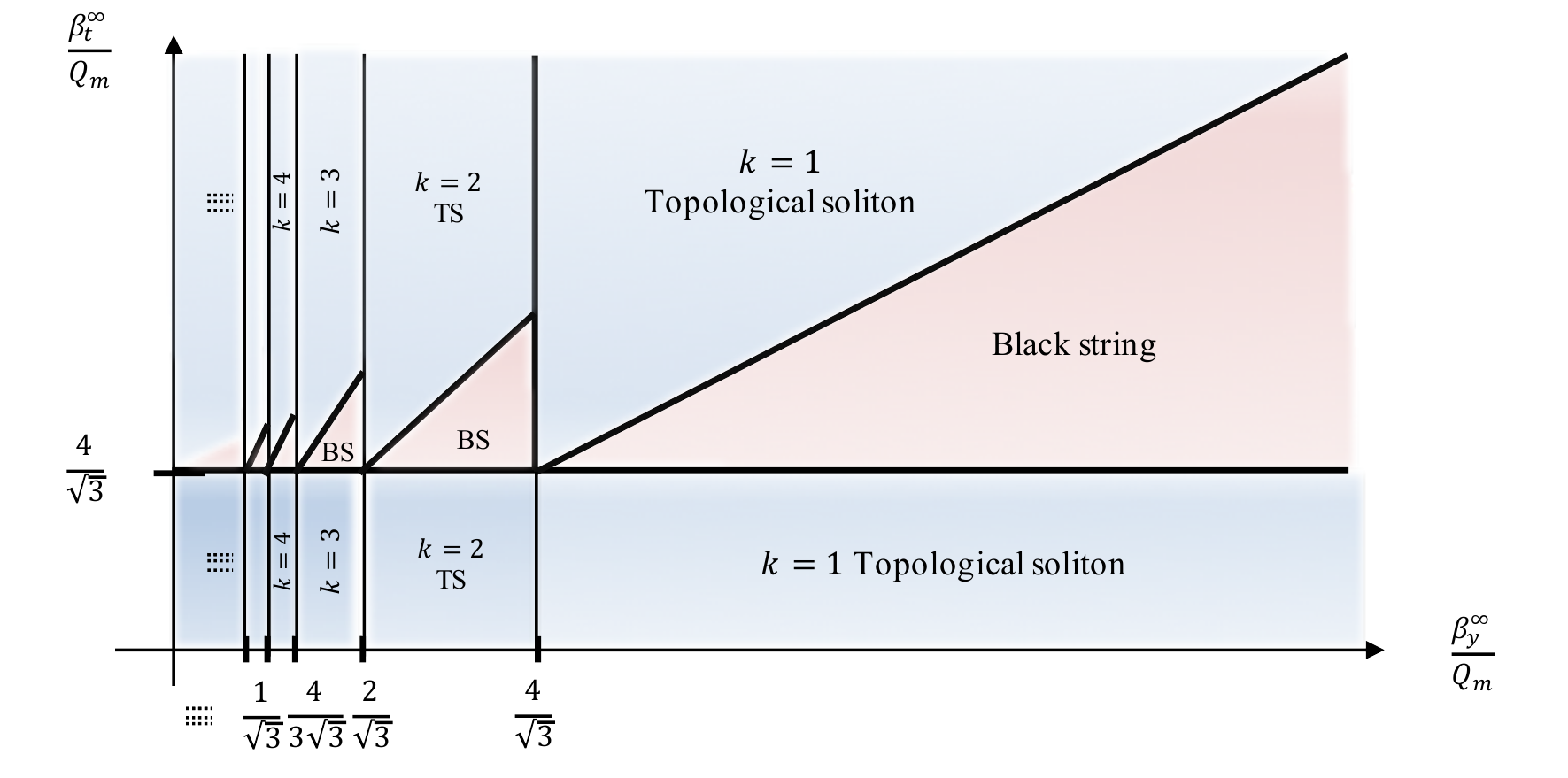}
\caption{The globally-stable phases between the black string and topological solitons.}
\label{fig:phaseBSTS}
\end{figure}

In \figref{fig:phaseBSTS}, we have summarized the global phase structure involving the locally-stable black string and topological soliton phases 
in the $(\frac{\beta_y^\infty}{Q_m}, \frac{\beta_t^\infty}{Q_m})$-plane. In the case where $\beta_t^\infty = k_\text{min} \beta_y^\infty$, 
the two locally-stable phases have the same free energy, i.e. $F_{\text{BS}} = F_{\text{TS}} $. Therefore, both phases can coexist and 
one can have a transition from one phase to the other, similar to a standard first order transition.

\subsection*{Stability and the extremal limit}

An interesting question to consider is the origin of these locally-stable phases for both the topological soliton and the black string.  This physics can be understood by considering the extremal black string and its relation to them.  Although the extremal system does not obey the boundary condition for the canonical ensemble, we can still consider the extremal limit for the off-shell free energy, i.e. $r_B = r_S$ at fixed $\beta_y^\infty$.  Whether we take the limit from the black string or from the topological soliton side, we obtain
\begin{equation}
F_{EBS} = \frac{2\pi^2\beta_y^\infty}{G_5} \frac{3r_E^2 + Q_m^2}{4r_E}, \qquad r_S = r_B = r_E.  
\end{equation}  This free energy is consistent with directly reducing the Euclidian action for the system while imposing the extremality condition.  Notice the free energy is positive and bounded from below unlike those of the black string and topological soliton.  It has a single extremum which is a global minimum and corresponds to a globally stable phase.  It satisfies:
\begin{equation}
Q^2_{m} = 3 r_E^2, \qquad F_{EBS}|_{\text{extremum}} = \frac{2\pi^2\beta_y^\infty}{G_5} \frac{\sqrt{3}|Q_m|}{2} .  
\end{equation}  Notice that the extremization does not fix $\beta_y^\infty$.  This is consistent with the fact we are free to fix the periodicity of the compact circle in this case. However, we have to understand this solution as existing at zero temperature.  

Since the extremal solution is globally stable, one can understand the stability of black string as a consequence of deformation away from the extremal point by turning on temperature.  Since the bubble solution is related to the black string by analytic continuation, we can similarly see the extremal solution as being responsible for the meta-stable phase of the topological soliton.  In figure \ref{fig:BSpot} the extremal solution can be associated to the boundary of the surface plot that is positive and increasing.  

\section{Conclusion and outlook} \label{sec:conclude}

In this paper, we have initiated a careful study of the thermodynamic/quantum-mechanical stability of black strings and topological solitons in a 
five-dimensional Einstein-Maxwell theory, with conserved magnetic and/or electric charges. We discuss, in particular, the locally-stable thermodynamic 
phases of the system, and address the question of meta-stability of the thermodynamic phases within the semi-classical regime. 
In addition, we address the question of global stability, and discuss the Hawking-Page transitions from 
one locally-stable phase to another. 

For the purposes of this paper, we work entirely in the canonical ensemble, where we put the five-dimensional theory in an Euclidean spherical 
cavity in $\BR^3 \times S^1_\tau \times S^1_y$, with $\tau$ and $y$ labeling the thermal circle and KK circle respectively.
The associated boundary conditions involve fixing the charge and the radius of the circles at the boundary. 
The gravitational path integral with these boundary conditions has two distinct types of real saddle points 
-- black strings and topological solitons with conical defects.
Using the standard tool-box of semi-classical gravity, we then investigate the thermodynamic phases for each type.

The principal tool for our analysis is the off-shell reduced action for the five-dimensional Einstein-Maxwell theory in the canonical ensemble,
and the associated free energy. We derive this reduced action in \Secref{sec:offShellAc}, and the answer, given in \eqref{RA-gen2}, is one of the main
results of our paper. The reduced action for a black string and a topological soliton can be read off from \eqref{RA-gen2} by appropriate 
analytic continuations of the five-dimensional Euclidean metric, and are given in \eqref{eq:EP-BS} and \eqref{eq:EP-TS} respectively. 

In each case, the off-shell action, which is a function of two variables, encodes a wealth of physics. 
Firstly, the extrema of this reduced action in each case give the precise thermodynamic phases of the system. 
Secondly, the local stability analysis of the extrema indicates whether a given thermodynamic phase is quantum-mechanically meta-stable. A locally-unstable extremum is identified with a gravitational instanton, which can mediate the decay of a locally-stable 
thermodynamic phase. The off-shell action can therefore be thought of as an effective potential 
for the system in the semi-classical regime.

In \Secref{sec:infbox}, we perform the aforementioned analysis for the black string and the topological soliton. 
In the derivation of the reduced action, the size of the spherical cavity plays the role of an IR regulator. 
For simplifying our presentation in \Secref{sec:infbox}, we set this regulator to infinity, which drastically 
simplifies the reduced action. From the analysis of the reduced action, we observe that there exists
a single locally-stable black string phase, provided the ensemble temperature is lower than a critical temperature 
which is determined by the charge. 
This black string, however, is meta-stable and can decay by the 
nucleation of a gravitational instanton, which turns out to be a larger black string. We compute the nucleation rate 
as a function of the boundary data, given in \eqref{eq:rateBS}, and note that the rate is severely 
suppressed at low temperatures, with the rest of the boundary data held fixed. 

For the topological solitons, we observe that there exists at most a single locally-stable phase 
for a given value of the orbifold parameter $k$. For any given boundary data, there exists a tower of 
such locally-stable phases, where the tower starts at a certain $k_{\text{min}}$ determined completely 
by the boundary data. Similar to the case of the black string, a locally-stable topological soliton, associated with 
the orbifold parameter $k \geq k_{\text{min}}$, turns out to be meta-stable, and can decay by the nucleation 
of a larger topological soliton with the same $k$. The nucleation rate, given in \eqref{eq:rateTS}, is severely 
suppressed in the limit where the size of the KK-circle is large, with other boundary data held fixed. In addition, 
there is a suppression factor of $k$, which implies that the topological soliton phase with higher $k$ is more stable 
against this decay. 

Given a set of locally-stable phases, one can address the question of global stability and possible Hawking-Page 
transitions. We find two distinct classes of transitions -- transitions between topological solitons with different orbifold 
parameters, and transitions between a topological soliton and a black string phase. For the first class of transitions, 
we find that a locally-stable topological soliton phase with $k > k_{\text{min}}$ can undergo a Hawking-Page transition 
to the globally-stable phase with $k = k_{\text{min}}$. For the second class of transitions,  the 
black string is the globally-stable phase in the high temperature regime, while the topological soliton phase 
with $k = k_{\text{min}}$ is globally-stable in the low temperature regime. There can also exist a regime of 
boundary data where both the topological soliton and the black string may coexist. The  global 
phase structure of the system is summarized in \figref{fig:phaseBSTS}.

For phenomenological reasons, it would be interesting to derive the complete expression for the decay rate of the meta-stable phases by computing the pre-factor $\Gamma_0$ in \eqref{eq:DecayRateGen}. The pre-factor is obtained from the computation of one-loop determinants around each instanton that mediates the decay. Moreover, to evaluate all possible channels of decay, one should also compute the rates that correspond to the Hawking-Page transitions between the black string and topological soliton phases.

There are a couple of special limits of the off-shell free energy that deserve some attention. 
Firstly, one can ask if the off-shell free energy that we compute reproduces the correct result 
in the extremal limit, although the limit might seem problematic given that it changes the 
topology of the boundary in the canonical ensemble. 
We directly confirm that our off-shell free energy does indeed give the correct 
answer for an extremal black string, and the free energy has a single global minimum, as expected.
This minimum corresponds to the globally-stable extremal black string phase. 
In addition, one observes that the meta-stable minimum in the non-extremal regime can be smoothly deformed to 
the globally-stable minimum in the extremal limit. Another interesting limit is the case of the vanishing magnetic charge $Q_m$. Here, the 
meta-stable minimum can be smoothly deformed to the locally-stable hot KK space-time.
We therefore observe that the meta-stable phase smoothly connects the hot KK space-time on the one hand 
and the extremal black string on the other, in the limit of zero charge and zero temperature respectively.

So far, we have only analyzed the thermodynamic phases that arise in the canonical ensemble in the 
limit of an infinite spherical cavity. We would like to conclude by mentioning some of the salient 
features of the analysis in a finite cavity. To begin with, the effective potential for the finite box 
must match the potential in the infinite box for the regime where the size of the black string or the 
topological soliton is small compared to the size of the box. In the regime closer to the boundary, the 
IR regulator adds a ``wall" to the effective potential for the infinite box depicted in \figref{fig:BSpot}, 
so that the potential is now bounded from below. 
This wall introduces a new minimum in the potential that corresponds to a locally-stable 
black string or topological soliton, which is almost as large as the size of the box. This is reminiscent of the four-dimensional Schwarzschild black hole 
in a finite spherical cavity in the canonical ensemble, which was discussed in \cite{York:1986it,Whiting:1988qr}. Since the potential is 
bounded from below, the locally-stable minima are not meta-stable any more but globally-stable, in contrast to the case of the infinite 
box.  This suggests that the charged geometries discussed in this paper, when embedded in AdS backgrounds, can be 
entirely stable. 

The finite cavity system leads to a much larger set of locally-stable phases, compared to the infinite cavity case, 
for both the black string and the topological soliton. This leads to a richer network of possible Hawking-Page transitions 
among the black string phases, the topological soliton phases as well as ones between a black string and topological soliton.
We expect to explore the physics of this system in a future work.

Finally, we emphasize that the formalism developed in this paper can be extended to study the 
quantum-mechanical stability for a large class of topological solitons, recently discussed in \cite{Bah:2020ogh,Bah:2020pdz,Bah:2021owp,Bah:2021rki,Heidmann:2021cms}. In particular, this requires constructing an off-shell reduced action in an 
axisymmetric set up, as opposed to the spherically symmetric situation that we have encountered in this work.
We expect to report on this problem in an upcoming paper.

\section*{Acknowledgments}
We are grateful to Kim Berghaus, Daniel Brennan, Melissa Diamond, David Kaplan, Zohar Komargodski, and Peter Weck for discussion. 
The work of IB, AD, and PH is supported in part by NSF grant PHY-2112699. The work of IB and AD is also supported in part by the Simons Collaboration on Global Categorical Symmetries.

\vspace{1cm}

\appendix
\leftline{\LARGE \bf Appendices}

\section{Classical stability of the black string}\label{app:GL-TS}

In this section, we investigate the classical stability of the black string (given by the metric \eqref{eq:metric5d} with $r_\text{S} > r_\text{B}$) 
by considering the following gravitational and gauge perturbations
\begin{equation}
g_{\mu \nu} \to g_{\mu\nu} + h_{\mu \nu} \,,\qquad F^{(x)} \to F^{(x)} + \delta F^{(x)} \,.
\end{equation}
The existence or absence of Gregory-Laflamme instabilities can be obtained by considering the threshold critical modes that correspond to the time-independent spherically-symmetric perturbations. Following the general prescription of \cite{Miyamoto:2007mh}, the spherically symmetric and static perturbations of the black string can be parametrized by the following metric ansatz:
\begin{align}
ds^2_5\= &  -\left(1-\frac{r_\text{S}}{r}\right)\, e^{2a(r,y)}\,dt^2 + \left(1-\frac{r_\text{B}}{r}\right)\, e^{2b(r,y)}\,\left(dy- \alpha(r,y)\,dr\right)^2 + \frac{e^{2c(r,y)}\,r^2 \, dr^2}{(r- r_\text{S})(r-r_\text{B})} \nn \\
&+ r^2\, e^{2d(r,y)}\, (d\theta^2 + \sin^2{\theta} \, d\phi^2), \nonumber\\
\equi & - f_t(r)\, e^{2a(r,y)}\,dt^2 + f_y(r)\,e^{2b(r,y)}\,(dy- \alpha(r,y)\,dr)^2 + \frac{e^{2c(r,y)}\, dr^2}{f_t(r)\,f_y(r)} \label{eq:perturbedmetric}  \\
& + r^2\, e^{2d(r,y)}\, (d\theta^2 + \sin^2{\theta} \, d\phi^2). \nonumber
\end{align} 
The gauge field strength $F^{(m)} = Q_m\,\sin\theta\,d\theta \wedge d\phi$ is still a solution in the above perturbed background, i.e. $d\star F^{(m)}=dF^{(m)}=0$. Therefore, one can show in the manner of \cite{Gregory:1994bj}, that one necessarily has $\delta F^{(m)} =0$ from the fact that $d\star \delta F^{(m)}=d \delta F^{(m)}=0$. As for $F^{(e)}$, $F^{(e)} = \frac{Q_e}{r^2}dr\wedge dt\wedge dy$ does not solve the Maxwell equations. However,  $e^{a+b+c-2 d}\frac{Q_e}{r^2}dr\wedge dt\wedge dy$ does. Therefore, if we redefine the perturbation scheme of $F^{(e)}$ such that $$F^{(e)} \to  e^{a+b+c-2 d} F^{(e)} + \bar{\delta} F^{(e)},$$
one can show that $\bar{\delta} F^{(e)}$ must be taken to zero to solve Maxwell and Bianchi equations as for the magnetic field perturbations. Under the perturbation, the gauge fields are therefore fixed to be
\begin{equation}
F^{(m)} \= Q_m\,\sin\theta\,d\theta \wedge d\phi\,,\qquad F^{(e)} \= e^{a+b+c-2 d}\frac{Q_e}{r^2}dr\wedge dt\wedge dy\,.
\end{equation}
Following \cite{Miyamoto:2007mh}, we shall impose the \textit{optimal gauge}:
\be
a(r,y) = b(r,y)=0,
\ee
with $c(r,y), d(r,y), \alpha(r,y)$ unrestricted. Einstein's equations for the gauge-fixed system are
\begin{align}
R_{\mu \nu}&~=~\frac{1}{2}\left( T_{\mu \nu} - \frac{1}{3} \,g_{\mu \nu} \, {T_\alpha}^\alpha\right)\,,\\
T_{\mu \nu} &~=~  {F^{(m)}}_{\mu \alpha}{{F^{(m)}}_{\nu}}^{ \alpha} - \frac{1}{4}\,g_{\mu \nu} {F^{(m)}}_{\alpha \beta}{F^{(m)}}^{\alpha \beta} +  \frac{1}{2} \, \left[{F^{(e)}}_{\mu \alpha \beta }{{F^{(e)}}_{\nu}}^{ \alpha\beta} - \frac{1}{6}\,g_{\mu \nu} {F^{(e)}}_{\alpha \beta \gamma}{F^{(e)}}^{\alpha \beta \gamma}\right]\,. \nonumber
\end{align}

The Einstein's equations lead to 5 independent equations (the $\theta\theta$ component and the $\phi\phi$ components are not independent) 
which involve $c,d,\alpha$ and their derivatives with respect to $r$ and $y$. The independent equations to the first order in the perturbations 
can be written as:
\begin{align}
& - \frac{r_\text{B} \,r_\text{S}\, f_t(r)}{r^4}\,\Big(c-2d \Big)  - \frac{r_\text{S}\, f_t(r)\,f_y(r)}{2r^2}\Big(\partial_r{c} - 2\partial_r{d}-\partial_y{\alpha}\Big) =0, \label{R-tt-1} \\
\nn \\
& \frac{r_\text{B} \,r_\text{S}\, f_y(r)}{r^4}\,\Big(c -2d \Big) + \frac{r_\text{B}\, f_t(r)\,f_y(r)}{2r^2}\,\Big(\partial_r{c} - 2\partial_r{d}\Big)
- \Big( \partial^2_y\, c + 2 \partial^2_y\, d \Big)\nn\\
 & - { f_t(r)\,f^2_y(r)}\, \partial_r \partial_y{\alpha} - \frac{ f_y(r)}{2r^3}(4r^2 - 2rr_\text{S} - rr_\text{B} - r_\text{B} r_\text{S}) \partial_y{\alpha} =0, \label{R-yy-1} \\
\nn  \\
 &  \frac{r_\text{B} \,r_\text{S}}{ f_t(r)\,f_y(r)\,r^4}\,\Big(c-2d \Big) + \frac{4r^2 -3rr_\text{S} -3rr_\text{B} + 2r_\text{S} r_\text{B}}{2r^3 \,f_y(r)\,f_t(r)}\Big(\partial_r{c} - 2\partial_r{d}\Big) 
- 2 \partial^2_r d \nn \\
& + \frac{4r_\text{S} r_\text{B} - r r_\text{S} - 3 rr_\text{B}}{2r^3 \,f_y(r)\,f_t(r)}\,\partial_y{\alpha} - \partial_r\partial_y{\alpha} 
- \frac{1}{f^2_y(r)\,f_t(r)}\, \partial^2_y{c} =0, \label{R-rr-1} \\
\nn \\
& \frac{3r_\text{B} - 2r}{r^2 f_y(r)}\,\partial_y{d} + \frac{-3r_\text{S} + 4r}{2r^2 f_t(r)}\,\partial_y{c} - 2 \partial_y \partial_r{d}=0, \label{R-yr-1} \\
\nn \\
& \frac{2 r_\text{B} \,r_\text{S}}{r^2}\,d+ 2 \Big(d - c\Big)\, \Big(-f_t(r) - f_y(r) + f_t(r)f_y(r) \Big) 
 + r f_t(r)f_y(r) \partial_r{c}  \label{R-thetatheta-1} \\ &+ \Big(\frac{-4r^2 + 3rr_\text{S} + 3rr_\text{B} - 2r_\text{S}r_\text{B}}{r} \Big)\, \partial_r{d} 
 - r^2 f_t(r)f_y(r) \partial^2_r{d} - r f_t(r)f_y(r) \partial_y{\alpha} - \frac{r^2}{f_y(r)} \partial^2_y {d}=0. \nonumber
\end{align}

The Einstein equations can be simplified in the following fashion. From the $tt$ and the $rr$ components, we get
\begin{align}
& \partial_r \Big( c- 2d \Big)= \partial_y \alpha - \frac{2r_\text{B}}{r^2\, f_y(r)}\, (c-2d), \\
& \partial_y \partial_r \alpha(r,y) = \frac{r_\text{B} r_\text{S} (c-2d)}{f_t(r)\,f_y(r)\, r^4} + \frac{4r^2 -3rr_\text{S} -3rr_\text{B} + 2r_\text{S} r_\text{B}}{2 r^3\,f_t(r)\,f_y(r)}\,\partial_r \Big( c- 2d \Big) \nn\\
& -2 \partial^2_r\, d + \frac{4r_\text{S} r_\text{B} - r r_\text{S} - 3 rr_\text{B}}{2r^3\, f_t(r)\,f_y(r)}\partial_y \alpha - \frac{1}{f_t(r)\,f^2_y(r)} \partial^2_y c. \nn
\end{align}
The equations can be used to eliminate the radial derivatives of $c$ and $\alpha$ from the rest of the equations. The reduced set of three 
equations (obtained by using the above constraints in the ${yy}$, ${yr}$, and ${\theta\theta}$ equations) are as follows:

\begin{align}
&  \partial^2_y\, d + \frac{f^2_y(r)\,(4r -3r_\text{S})}{2r^2}\,\partial_y\alpha - \frac{f_y(r)\,(4r -3r_\text{S})\,r_\text{B}}{2r^4}\,(c-2d)=f_t(r)\,f^2_y(r)\, \partial^2_r\, d, \nn \\
&  \frac{3r_\text{B} - 2r}{r^2 f_y(r)}\,\partial_y{d} + \frac{-3r_\text{S} + 4r}{2r^2 f_t(r)}\,\partial_y{c} - 2 \partial_y \partial_r{d}=0, \\
& -(2r-r_\text{S} -r_\text{B})\,\partial_r{d} + 2(c-d) - \frac{2r_\text{B}}{r}\,(c-2d)= r^2 f_t(r)f_y(r) \partial^2_r{d} + \frac{r^2}{f_y(t)} \partial^2_y {d}. \nn
\end{align}

Note that, in the reduced set of 3 equations, we have radial derivatives of the function $d(r,y)$ but not of $c(r,y)$ and $\alpha(r,y)$. 
Expanding $d, c, \alpha$ as Fourier series along the periodic direction $y$, one can therefore use two of the three equations as algebraic equations 
to eliminate $c$ and $\alpha$, leaving behind a single ordinary differential equation for $d$ (a given Fourier mode of $d$) as a function of $r$.\\

More precisely, one can substitute in the above equations:
\begin{align}
& c(r,y)= c(r)\,\cos{ky} , \quad d(r,y)=d(r)\,\cos{ky}, \\ \label{eq:modes}
& \alpha(r,y)= - \alpha(r) \,k\, \sin{ky},
\end{align}
where $k$ is the GL critical wavenumber to be determined. With this substitution, the non-trivial equation in $r$ is given as:
\be
- A(r)\, d''(r) + B(r)\, d'(r)+ V(r)\,d = 0\,,
\ee
where the functions $A(r)$ and $B(r)$ are given as:
\begin{align}
A(r)=f_t(r)\,f^2_y(r), \quad B(r)= \frac{f_y(r)}{r^2}\, \Big(-(2r-r_\text{S} -r_\text{B}) + \frac{8r^2\,f_t(r)\,f_y(r)}{4r-3r_\text{S}} \Big), 
\end{align}
and the potential $V(r)$ is given as:
\be \label{V-final}
\boxed{V(r)= k^2+ \frac{2( 2r_\text{B}-r_\text{S})\,(r-r_\text{B})}{(4r-3r_\text{S})\,r^3} .}
\ee
Note that on rescaling $r \to r/r_\text{S}$ and defining $q=r_\text{B}/r_\text{S}$, this potential reproduces equation (15) of \cite{Miyamoto:2007mh}.\\

The other two equations express $c$ and $\alpha$ in terms of $d$, i.e.
\begin{align}
& c(r) = \frac{2r^2 f_t(r)}{4r-3r_\text{S}}\,\Big(2d'(r) - \frac{3r_\text{B}-2r}{r^2 f_y(r)} d(r) \Big),\\
& -\frac{f^2_y(r)\,(4r -3r_\text{S})}{2r^2}\,k^2\,\alpha(r) = \Big(\frac{2f_y(r)f_t(r)}{r^2} +B(r) \Big)\,d'(r) \nn \\
& \hspace{3.3cm}+ \Big(-\frac{r_\text{B}\, f_t(r)\, (3r_\text{B} -2r)}{r^4} - \frac{f_y(r)\,(4r-3r_\text{S})r_\text{B}}{r^4} + V(r)+ 2 k^2 \Big)\,d(r). \nn
\end{align}

Given \eqref{V-final}, one can check that for a background corresponding to a black string (i.e. $r > r_\text{S} > r_\text{B}>0$), the potential is strictly positive for all $r$ if and only if 
\be
q = \frac{r_\text{B}}{r_\text{S}} > \frac{1}{2}.
\ee
Therefore, there is no solution $d(r)$ that is vanishing asymptotically and at the horizon that is different from 0. However, when $q<\frac{1}{2}$, the second term in the potential is negative and one can play with $k$ in order to have a positive potential at the horizon and at the asymptotics with a negative region in between. For such configuration, regular static solutions exist and highlight the presence of critical Gregory-Laflamme instabilities. Thus, the black string solutions are classically stable in the regime $\frac{r_\text{B}}{r_\text{S}} > \frac{1}{2}.$

In \cite{Stotyn:2011tv}, it has been argued that since a black string with $\frac{r_\text{B}}{r_\text{S}} < \frac{1}{2}$ is unstable, the topological soliton with $\frac{r_\text{S}}{r_\text{B}} < \frac{1}{2}$ is also unstable by Wick rotation symmetry. We can demonstrate this fact more clearly from the above computation for the black string. 
The unstable mode $d(r)$, given in \eqref{eq:modes}, depends on the KK-circle coordinate by a factor $\cos{ky}$. Under the double analytic 
continuation: $t \to iy$ and $y \to it$ with the parameters $r_\text{S} \leftrightarrow r_\text{B}$, this mode is mapped to a time-dependent growing mode for the topological soliton proportional to $\cosh{kt}$. This indicates that the topological soliton has a classical instability in the regime $\frac{r_\text{S}}{r_\text{B}} < \frac{1}{2}$.

\section{Thermodynamics of five-dimensional Einstein-Maxwell theory with an electric field}\label{app:electric charge}

We consider the Einstein-Maxwell theory given by \eqref{eq:Action5d} with a fixed magnetic and electric charges on a manifold $\CM$ with boundary $\Sigma$, in the canonical ensemble. 
We choose $\Sigma$ to have the topology of $S^1 \times S^1 \times S^2$, where the first $S^1$ is the thermal circle and 
the second $S^1$ is the KK circle.

The canonical ensemble is defined by putting the system inside a spherical cavity of radius $R_b$ and fixing the radius $\beta_t$ of the thermal circle as well as the radius $\beta_y$ of the KK circle at the boundary of the cavity. In addition, we fix the magnetic and electric charges, $Q_m$ and $Q_e$, enclosed 
in the cavity: 
\begin{equation}
Q_m \= \frac{1}{4\pi} \int_{S^2_{\Sigma}} F^{(m)}\,,\qquad Q_e \= \frac{-i}{4\pi} \int_{S^2_{\Sigma}} \star_5 F^{(e)}\,.
\end{equation}
The classical action of the system is given as:
\begin{equation}
\begin{split}
I &\=-\frac{1}{16 \pi G_{5}} \int_{\mathcal{M}} d^{5} x \sqrt{g}\left(R-\frac{1}{4} F^{(m)}_{a b} F^{(m)\,a b} -\frac{1}{12} F^{(e)}_{a b c} F^{(e)\,a b c}\right)+I_\text{boundary},\\
I_\text{boundary} &\= - \frac{1}{8 \pi G_{5}} \int_{\partial \cM} d^4 x \,\sqrt{h}\,\left( K - K_0 \+ \frac{1}{2} n_a F^{(m)\,ab} A^{(m)}_b \+  \frac{1}{4} n_a F^{(e)\,abc} A^{(e)}_{bc}\right)\,,
\end{split}
\label{eq:EucAction}
\end{equation}
where the last terms are the standard boundary terms for the Maxwell fields in the canonical ensemble, with $n_a$ being the normal unit vector 
to the boundary $\Sigma$.\\

Following the general philosophy outlined in \Secref{sec:review}, the first step is to list the saddle points of the theory, i.e. 
solutions of Einstein's equations consistent with the boundary conditions. 
The two types of saddle points for this theory are identical to the magnetic case and correspond to an Euclidean black string and topological soliton given by \eqref{BSmet-1} and \eqref{SBmet-1} with an additional electric field,
\begin{equation}
F^{(e)} \= \frac{i \,Q_e}{r^2}\,d\tau \wedge dr \wedge dy\,.
\end{equation}
The regularity conditions for both solutions are identical to the magnetic case detailed in section \ref{sec:saddle} by replacing all $Q_m^2$ by the total charge $\cQ^2\equi Q_m^2+Q_e^2$.

\subsection{Off-shell Reduced action for the Euclidean theory}
 
The analysis of the free energy in the presence of an electric charge is then identical to the exercise in \Secref{sec:offShellAc},
once we replace the magnetic charge by the effective charge $\cQ^2 \equi Q_m^2+Q_e^2$. 
Let us check this explicitly. We consider the following ansatz:
\begin{align}
d s^{2}&=U^{2}(\rho) d x_1^{2}+V^{2}(\rho) d x_2^{2}+\frac{1}{W^{2}(\rho) V^{2}(\rho)} d \rho^{2}+r^{2}(\rho)\, d \Omega_{2}^{2},\\
F^{(e)} &= dA_t(\rho)\wedge dt\wedge dy,\qquad F^{(m)} = dA_\phi(\theta)\wedge d\phi,
\end{align}
where the two circle directions are labelled by $x_i$. The boundary is at $r=R_b$ given by $\rho=1$. 
In addition, we assume that the function $U$ has a zero at a certain $r=r_1$,  given by $\rho=0$, while the function $V(r)$ has a zero at a different $r=r_2$. We assume $ r_1 > r_2$, so that the $x_2$-circle does not 
shrink in the manifold $\cM=\{0\leq \rho\leq 1\}$. The
$x_i$-circle has a period $2\pi \beta_i$ at the boundary $r=R_b$. Note that we are not identifying any of the circle directions as the thermal circle or the KK circle at this point.

\begin{itemize}
\item[1.]The boundary condition implies that: 
$$
\begin{aligned}
& \beta_1 = \frac{1}{2\pi}\,\int^{2\pi}_{0} U(1) dx_1 =  U(1),\quad \beta_2= \frac{1}{2\pi}\,\int_{0}^{2 \pi} V(1) d x_2=V(1),\\
&Q_m \= \frac{1}{4\pi} \int_{S_\Sigma^2} F^{(m)}\bigl|_{\rho=1}\,,\qquad Q_e \= \frac{-i}{4\pi} \int_{S^2_{\Sigma}} \star_5 F^{(e)} \bigl|_{\rho=1}\,.
\end{aligned}
$$
\item[2.] Regularity of the metric at the origin implies that :
$$
[W(\rho)\, V(\rho)\, U'(\rho)]_{\rho=0} =k^{-1}.
$$
\item[3.] We impose the Hamiltonian constraint and the Maxwell equation for the electric gauge field (the magnetic one will not relevant for the derivation):
\begin{equation}
\begin{split}
&G^\tau_{\,\,\tau}\= \frac{1}{8}\left(\frac{1}{3} \left|F^{(e)}\right|^2-\left|F^{(m)}\right|^2 \right)\,,\\
&\partial_\rho \left( \frac{r^2 W A_t'}{U} \right) \=0 \qquad \Leftrightarrow \qquad \frac{r^2 W A_t'}{U} \= \text{cst}.
\end{split}
\end{equation}
By using the condition on the charge we find $\text{cst}=-i Q_e$.
\end{itemize}
We now evaluate the different parts of the action \eqref{eq:EucAction}, using the constraint above,
\begin{equation}
\begin{split}
&-\frac{\sqrt{g}}{16\pi G_5} \left(R-\frac{1}{4} \left|F^{(m)}\right|^2- \frac{1}{12}\left|F^{(e)}\right|^2\right) \= \frac{\sin \theta}{16\pi G_5} \left[2 \partial_\rho\left(r^2 W V^2 U' \right) - i Q_e\,A_t' \right],\\
& -\frac{\sqrt{h}}{8\pi G_5}\left( K-K_0 \right) \bigl|_{\rho=1}\= \frac{V(1)\,\sin \theta}{8\pi G_5} \left[2 R_b U(1)  -W(1)\,\partial_\rho\left( r^2 U V\right)\bigl|_{\rho=1} \right],\\
& -\frac{\sqrt{h}}{16\pi G_5}\left(  n_a F^{(m)\,ab} A^{(m)}_b \+  \frac{1}{2} n_a F^{(e)\,abc} A^{(e)}_{bc}\right)\Bigl|_{\rho=1}  \=  \frac{\sin \theta}{16\pi G_5}\,i\,Q_e\,A_t(1).
\end{split}
\end{equation}

Gathering all the terms, and integrating over the angular coordinates and the radial coordinate, the reduced action is
\begin{equation}
I\= \frac{2\pi^2}{G_5}\left[ U(1) V(1)\left(2R_b-W(1)\, (r V)'(1)\right)-\frac{r_1^2}{k} \,V(1) \right],
\end{equation}
and we manifestly retrieve the same action \eqref{RA-gen0}.



\bibliographystyle{utphys}      

\bibliography{microstates}       


\end{document}